\documentclass[sn-mathphys]{sn-jnl}
\usepackage[utf8]{inputenc}
\usepackage[T1]{fontenc}
\usepackage{amsmath}
\usepackage{amsfonts}
\usepackage{amssymb}
\usepackage[version=4]{mhchem}
\usepackage{stmaryrd}
\usepackage{graphicx}
\usepackage[export]{adjustbox}
\usepackage{csquotes}
\usepackage{hyperref}
\usepackage{color}
\usepackage{amsthm}
\usepackage{tikz}
\usetikzlibrary{decorations.pathreplacing} 
\usetikzlibrary{arrows.meta} 
\usepackage{orcidlink}

\newcommand{\HRule}[1]{\rule{\linewidth}{#1}}

\title{ \normalsize \textsc{}
		\HRule{2pt} \\
  		\LARGE \textbf{Characterization of exotic matter\\
in $\mathcal{PT}$-symmetric wormholes} 
		\HRule{2.0pt}}

\newcommand{\runhead}{\textit{Characterization of exotic matter in $\mathcal{PT}$-symmetric wormholes}}

\jyear{\the\year}

\begin{document}

\title[Characterization of exotic matter in $\mathcal{PT}$-symmetric wormholes]
{Characterization of exotic matter in $\mathcal{PT}$-symmetric wormholes}

\author{\small Hicham Zejli%
  \thanks{ORCID: \href{https://orcid.org/0009-0006-8886-7101}%
  {0009-0006-8886-7101}}}

\affil{\small \itshape Independent Researcher, France}

\abstract{
In our previous work [H. Zejli, Int. J. Mod. Phys. D 34, 2550052 (2025), arXiv:2508.00035], we introduced a $\mathcal{P}\mathcal{T}$-symmetric wormhole model based on a bimetric geometry, capable of generating closed timelike curves (CTCs). In this paper, we extend the analysis to the null hypersurface at the throat of this modified Einstein-Rosen bridge, where two regular Eddington-Finkelstein metrics render the geometry traversable. Using the Barrabès-Israël formalism in Poisson’s reformulation, we evaluate the null shell’s surface stress-energy tensor $S^{\alpha\beta}$ from the jump of the transverse curvature, revealing a violation of the null energy condition: a lightlike membrane of exotic matter with negative surface energy density and positive tangential pressure. This exotic fluid acts as a repulsive source stabilizing the throat, ensuring consistency with the Einstein field equations, including conservation laws on the shell. Beyond the local characterization, we outline potential observational signatures: (i) gravitational-wave echoes from the photon-sphere cavity; (ii) horizon-scale imaging with duplicated and through-throat photon rings, and non-Kerr asymmetries; (iii) quantum effects such as $\mathcal{PT}$-induced frequency pairing with possible QNM doublets and partial suppression of vacuum flux at the throat; and (iv) a relic cosmological population yielding an effective $\Lambda_{\rm eff}$ and seeding voids. Compared with timelike thin-shell constructions, our approach is based on a null junction interpreted as a lightlike membrane, combined with $\mathcal{PT}$ symmetry, providing a distinct route to traversability and clarifying the conditions under which CTCs can arise in a self-consistent framework.
}


\maketitle

\newpage

\setcounter{tocdepth}{2}
\markboth{\textit{Contents}}{\textit{Contents}}
\tableofcontents
\markboth{\textit{Contents}}{\textit{Contents}}
\clearpage
\markboth{\runhead}{\runhead}

\section{Introduction}
\label{sec:intro}
Traversable wormholes, solutions to the equations of general relativity, have garnered significant interest as hypothetical passages connecting distant regions of spacetime. Their theoretical existence, however, relies on stringent physical conditions, notably the presence of exotic matter to keep their throat open. In previous work \cite{koiran2024,Zejli2025}, we proposed an innovative model of a modified Einstein-Rosen bridge, rendered unidirectionally traversable through a bimetric\footnote{In the context of~\cite{koiran2024}, ``bimetric'' refers to a single spacetime described by two metric forms, where the metric is continuous across the throat but its first derivatives exhibit a discontinuity, compensated by a lightlike membrane of exotic matter to satisfy the Einstein field equations~\cite{guendelman2010,guendelman2016einstein,koiran2024}.} geometry defined by two regular metrics, $g^{(+)}$ and $g^{(-)}$, and characterized by $\mathcal{PT}$ symmetry (parity and time reversal). This model, formulated in Eddington-Finkelstein coordinates, eliminates coordinate singularities at the throat $r = \alpha$ \footnote{Where $\alpha = 2M$ represents the Schwarzschild radius for a mass $M$.} and introduces a lightlike membrane of exotic matter at the junction, thereby satisfying Einstein’s field equations, as seen in other traversable wormhole models \cite{HochbergVisser1997, Morris1988, visser1995}. By coupling two such wormholes, we demonstrated the possibility of generating closed timelike curves (CTCs), consistent with Novikov’s self-consistency principle, thus opening new perspectives on causality in traversable geometries.\\

In this study, we extend the analysis of the hypersurface located at the throat of this $\mathcal{PT}$-symmetric wormhole, focusing on the geometric and physical structure of the null hypersurface $\Sigma$ at $r = \alpha$. This hypersurface separates two spacetime regions described by the following metrics:
\begin{itemize}
    \item Incoming metric \( g^{(+)} \) :
    \begin{equation}\label{eq_line_element_incoming}
    \mathrm{d}s_+^2 = \left(1 - \frac{\alpha}{r}\right) \mathrm{d}t^2 - \left(1 + \frac{\alpha}{r}\right) \mathrm{d}r^2 - \frac{2\alpha}{r} \mathrm{d}r \, \mathrm{d}t - r^2 \left(\mathrm{d}\theta^2 + \sin^2\theta \, \mathrm{d}\phi^2\right)
    \end{equation}
    \item Outgoing metric \( g^{(-)} \) :
    \begin{equation}\label{eq_line_element_outgoing}
    \mathrm{d}s_-^2 = \left(1 - \frac{\alpha}{r}\right) \mathrm{d}t^2 - \left(1 + \frac{\alpha}{r}\right) \mathrm{d}r^2 + \frac{2\alpha}{r} \mathrm{d}r \, \mathrm{d}t - r^2 \left(\mathrm{d}\theta^2 + \sin^2\theta \, \mathrm{d}\phi^2\right)
    \end{equation}
\end{itemize}

These metrics are related by $\mathcal{PT}$ symmetry, which combines time reversal ($t \rightarrow -t$) and spatial inversion ($\vec{x} \rightarrow -\vec{x}$), ensuring a coherent structure between the two regions of the traversable spacetime\footnote{Geometrized units $G=c=1$ and signature $(+---)$ are used throughout \cite{koiran2024,Zejli2025}.}\footnote{Locally, the mixed term can be removed on a single Eddington–Finkelstein sheet by a time redefinition $t\!\to\! t\pm f(r)$, yielding a diagonal form. In the present bimetric construction, however, the two sheets are time-reversed copies (see the pair (17)–(18)), so the diagonalizing transformations come with opposite signs on $\mathcal{M}_\pm$. A single global change of coordinates that kills $g_{tr}$ on both sheets is therefore incompatible with the $\mathcal{PT}$-symmetric gluing at $r=\alpha$ and would also spoil the finite–coordinate–time crossing property. We thus retain $g_{tr}\neq 0$ to preserve the global structure (see \cite[Sec.~3.2]{koiran2021} and \cite[Sec.~4]{koiran2024}.}. They are continuous at $r = \alpha$, but their first derivatives exhibit a discontinuity, giving rise to a delta-type singularity typical of lightlike hypersurfaces\footnote{Eqs.~\ref{eq_line_element_incoming}–\ref{eq_line_element_outgoing} are the ingoing/outgoing Eddington–Finkelstein forms of the Schwarzschild exterior obtained by a radially dependent time redefinition $t\mapsto t\pm f(r)$ (no change of $r$, see Eqs. (10) and (12) in \cite{koiran2024}). Hence they are not ad hoc but inherit the standard physical content of the Schwarzschild vacuum solution. The non-smooth chart $r=\alpha+|\eta|$ at the throat produces the null shell and induces the natural $\mathcal{PT}$ pairing of the two sheets (see Eq. (13) and the bimetric pair (17)–(18) in \cite{koiran2024}). This is the standard Barrabès–Israël picture for null shells \cite{Barrabes1991,Poisson2002Reformulation}.}. To address this discontinuity, we employ the Barrabès-Israël formalism \cite{Barrabes1991,Poisson2002Reformulation}, ideally suited for null hypersurfaces. This framework, which generalizes Israël’s foundational work \cite{israel1966} on thin shells, enables the calculation of the surface stress-energy tensor $S^{\alpha\beta}$ at the throat. This analysis also draws on Poisson’s reformulation \cite{Poisson2002Reformulation} of the Barrabès-Israël formalism for null hypersurfaces, which provides complementary insights into energy conditions in the context of wormholes and similar geometries. As we will demonstrate, our results confirm a violation of the null energy condition (NEC), marked by a negative energy density, underscoring the necessity of exotic matter to ensure traversability.\\

The main objectives of this work are as follows:
\begin{itemize}
    \item To study the null hypersurface $\Sigma$ at the throat of the wormhole using the Barrabès-Israël formalism, in order to characterize its geometric and physical properties.
    \item To analyze and perform the invariant null–shell decomposition of the surface stress-energy tensor $S^{\alpha\beta}$ according to the Barrabès-Israël formalism \cite{Barrabes1991} reformulated by Poisson \cite{Poisson2002Reformulation}, to reveal the characteristics of the exotic matter, notably a negative surface energy density and positive tangential pressure, essential for stabilizing the traversable geometry.
    \item To explore the implications of these properties for the dynamic stability of the throat and their role in maintaining the traversability of the wormhole.
    \item To connect these results to the closed timelike curves (CTCs) studied in our previous work \cite{Zejli2025}, deepening the understanding of the throat’s role in the formation of CTCs.
\item To explore a speculative interpretation of the exotic matter within bimetric frameworks, 
potentially offering insights into its nature and cosmological implications.
\item {To translate the model into falsifiable observational signatures across multiple channels:}
(i) \emph{gravitational waves} : late-time \emph{“echoes”} in post-merger signals caused by partial reflections between the light ring and the throat;
(ii) \emph{horizon-scale imaging} : duplicated or \emph{“through-throat”} photon rings and brightness asymmetries that are {absent in Kerr} (where images show a single Kerr-like ring);
(iii) \emph{quantum sector} : $\mathcal{PT}$-induced frequency pairing $\omega\!\leftrightarrow\!-\omega$, partial cancellations of vacuum flux at the throat, and symmetric (possibly complex-conjugate) quasinormal-mode (QNM) doublets;
(iv) \emph{cosmology} : a relic population acting as an effective $\Lambda_{\rm eff}$ and seeding cosmic voids, with constraints from {non}-detections (no resolvable echoes and EHT images consistent with a single Kerr-like ring).
\end{itemize}

The structure of the article is as follows:
Section~\ref{sec:section2} characterizes the null hypersurface $\Sigma$ at the throat ($r = \alpha$), defining its geometry and demonstrating its null nature via the normal vector. Section \ref{sec:section3} establishes a suitable vector basis on $\Sigma$, followed by the calculation of the transverse curvature tensor in Section \ref{sec:section4}. Sections \ref{sec:section5} and \ref{sec:tenseur-energie-impulsion} address the curvature jump and the determination of the surface energy density and the pressure through the invariant null–shell decomposition of the surface stress-energy tensor $S^{\alpha\beta}$, according to the Barrabès-Israël formalism \cite{Barrabes1991} reformulated by Poisson \cite{Poisson2002Reformulation}, respectively. Section \ref{sec:nec_violation} analyzes the violation of the NEC, while Section \ref{sec:pt_coherence} provides a detailed physical interpretation of $S^{\alpha\beta}$, including its implications for the stability of the throat. Section \ref{sec:section9} explores a bimetric perspective for interpreting the exotic matter. {Section~\ref{sec:predictions}} then presents the model’s testable {predictions and observational signatures} (GW echoes, horizon-scale imaging, quantum spectral patterns, and cosmological effects), with technical complements in {Appendix~\ref{sec:appendix_Lcav}} (echo cavity length) and {Appendix~\ref{sec:toy2D}} (2D DFU toy model). Finally, Section~\ref{sec:section10} concludes with perspectives and constraints arising from current and forthcoming observations.

\section{Characterization of the null hypersurface at the wormhole throat}
\label{sec:section2}

In this section, we define the hypersurface \(\Sigma\) located at the wormhole throat, at \( r = \alpha \), and demonstrate that it's a null hypersurface by computing its normal vector. Unlike the classical definition of the throat as a hypersurface of minimal area in a constant-time spatial slice, as proposed by Hochberg and Visser~\cite{HochbergVisser1997}, our approach adopts a null hypersurface to reflect the lightlike nature of the junction in a $\mathcal{P}\mathcal{T}$-symmetric bimetric model. This hypersurface plays a key role in separating two spacetime regions endowed with distinct regular metrics, denoted \( g^{(+)} \) and \( g^{(-)} \), which correspond to the incoming and outgoing sides of the wormhole, respectively.

\subsection{Definition of the hypersurface \(\Sigma\)}

The hypersurface \(\Sigma\) is defined by the following condition:
\begin{equation}
F(r) = r - \alpha = 0,
\end{equation}
where \(\alpha\) represents the radius of the throat. The covariant normal vector to this hypersurface, \( n_\mu \), is given by the gradient of the function \( F \), namely:
\begin{equation}
n_\mu = \partial_\mu F.
\end{equation}
We compute the partial derivatives of \( F \) with respect to the coordinates \( x^\mu = (t, r, \theta, \phi) \):
\begin{itemize}
    \item \(\partial_t F = 0\), since \( F \) doesn't depend on \( t \),
    \item \(\partial_r F = \frac{\partial}{\partial r} (r - \alpha) = 1\),
    \item \(\partial_\theta F = 0\) (no dependence on \( \theta \)),
    \item \(\partial_\phi F = 0\) (no dependence on \( \phi \)).
\end{itemize}
Thus, the covariant normal vector is:
\begin{equation}
n_\mu = (0, 1, 0, 0),
\end{equation}
indicating that it's oriented in the radial direction \( r \)\footnote{only the component \( \partial_r F = 1 \) is non-zero.}.

\subsection{Verification of the null nature of \(\Sigma\)}

To establish that \(\Sigma\) is a null hypersurface, we must show that the norm of its normal vector is zero, i.e.:
\begin{equation}
n^\mu n_\mu = g^{\mu\nu} n_\mu n_\nu = 0,
\end{equation}
where \( g^{\mu\nu} \) is the inverse metric evaluated at \( r = \alpha \). Since \(\Sigma\) separates two regions with metrics \( g^{(+)} \) and \( g^{(-)} \), we must examine the consistency of this property on both sides of the junction. We begin with the metric \( g^{(+)} \) on the incoming side and then confirm that the result is similar for \( g^{(-)} \).

\subsubsection{Calculation of the norm of the normal vector \( n_\mu \) with the metric \( g_{\mu\nu}^{(+)} \)}

At \( r = \alpha \), the non-zero components of the metric \( g_{\mu\nu}^{(+)} \) are given by the line element \ref{eq_line_element_incoming}. The relevant components associated with this geometry are:
\begin{equation}\label{eq_gplus_covar}
g_{tt}^{(+)} = 0, \quad g_{rr}^{(+)} = -2, \quad g_{tr}^{(+)} = g_{rt}^{(+)} = -1,
\end{equation}
along with the angular components \( g_{\theta\theta}^{(+)} = -\alpha^2 \) and \( g_{\phi\phi}^{(+)} = -\alpha^2 \sin^2 \theta \). The metric matrix at \( r = \alpha \) is thus:
\begin{equation}
g_{\mu\nu}^{(+)} = \begin{pmatrix}
0 & -1 & 0 & 0 \\
-1 & -2 & 0 & 0 \\
0 & 0 & -\alpha^2 & 0 \\
0 & 0 & 0 & -\alpha^2 \sin^2 \theta
\end{pmatrix}.
\end{equation}
To compute the inverse metric \( g_{(+)}^{\mu\nu} \), we focus on the \( (t, r) \) block, as the angular components don't contribute to the product \( n^\mu n_\mu \) with \( n_\mu = (0, 1, 0, 0) \). The \( (t, r) \) sub-block is:
\begin{equation}
g_{(t,r)}^{(+)} = \begin{pmatrix}
0 & -1 \\
-1 & -2
\end{pmatrix}.
\end{equation}
Its determinant is:
\begin{equation}
\det g_{(t,r)}^{(+)} = -1.
\end{equation}
The inverse of this sub-block is calculated as follows:
\begin{equation}
g_{(+)}^{\mu\nu}\Big|_{(t,r)} = \frac{1}{\det g_{(t,r)}^{(+)}} \begin{pmatrix}
-2 & 1 \\
1 & 0
\end{pmatrix} = \begin{pmatrix}
2 & -1 \\
-1 & 0
\end{pmatrix}.
\end{equation}
Thus, the relevant components of the inverse metric are:
\begin{equation}\label{eq_gplus_contra}
g_{(+)}^{tt} = 2, \quad g_{(+)}^{tr} = g_{(+)}^{rt} = -1, \quad g_{(+)}^{rr} = 0.
\end{equation}

Hence:
\begin{equation}
n^\mu n_\mu = g_{(+)}^{\mu\nu} n_\mu n_\nu = g_{(+)}^{rr} n_r n_r = 0
\end{equation}

This confirms that \(\Sigma\) is a null hypersurface on the side of the metric \( g^{(+)} \).

\subsubsection{Consistency with the metric \( g_{\mu\nu}^{(-)} \)}

For the outgoing side, the metric tensor \( g_{\mu\nu}^{(-)} \) derived from the line element \ref{eq_line_element_outgoing} is given by:
\begin{equation}
g_{\mu\nu}^{(-)} = \begin{pmatrix} 
0 & 1 & 0 & 0 \\ 
1 & -2 & 0 & 0 \\ 
0 & 0 & -\alpha^2 & 0 \\
0 & 0 & 0 & -\alpha^2 \sin^2 \theta
\end{pmatrix}.
\end{equation}

Thus, at \( r = \alpha \), we obtain:
\begin{equation}\label{eq_gmoins_covar}
g_{tt}^{(-)} = 0 , \quad g_{rr}^{(-)} = -2, \quad g_{tr}^{(-)} = 1.
\end{equation}

The \( (t, r) \) sub-block becomes:
\begin{equation}
g_{(t,r)}^{(-)} = \begin{pmatrix}
0 & 1 \\
1 & -2
\end{pmatrix},
\end{equation}
with an identical determinant:
\begin{equation}
\det g_{(t,r)}^{(-)} = -1.
\end{equation}
The inverse is then:
\begin{equation}
g_{(-)}^{\mu\nu}\Big|_{(t,r)} = \begin{pmatrix}
2 & 1 \\
1 & 0
\end{pmatrix},
\end{equation}
yielding:
\begin{equation}\label{eq_gmoins_contra}
g_{(-)}^{tt} = 2, \quad g_{(-)}^{tr} = g_{(-)}^{rt} = 1, \quad g_{(-)}^{rr} = 0.
\end{equation}

Considering \( n_\mu = (0, 1, 0, 0) \), we obtain:
\begin{equation}
n^\mu n_\mu = g_{(-)}^{\mu\nu} n_\mu n_\nu = g_{(-)}^{rr} = 0
\end{equation}

The null property of the hypersurface \(\Sigma\) is thus also verified for \( g^{(-)} \).\\

The hypersurface \(\Sigma\) at \( r = \alpha \) is therefore a null hypersurface for both metrics \( g^{(+)} \) and \( g^{(-)} \). This result is consistent with the role of \(\Sigma\) as a lightlike junction hypersurface in the \(\mathcal{PT}\)-symmetric wormhole model. This property reflects its lightlike nature, typical of throat descriptions in traversable geometries.

\subsubsection{Continuity of the induced metric on $\Sigma$}
Let $q_{ab}:=g_{\mu\nu}\,e_a^{\mu}e_b^{\nu}$ denote the (degenerate) metric induced on $\Sigma$
by the basis $\{k^\mu,e^\mu_\theta,e^\mu_\phi\}$. \\

On $r=\alpha$ one has
\begin{equation}
q_{tt}=0,\qquad q_{\theta\theta}=-\alpha^2,\qquad q_{\phi\phi}=-\alpha^2\sin^2\theta,
\end{equation}
identical on the $g^{(+)}$ and $g^{(-)}$ sides. Thus the Barrabès-Israël continuity requirement on the induced geometry is satisfied. In particular, once this holds, the junction data are entirely carried by the transverse curvature $C_{ab}$. Its jump $[C_{ab}]$ (defined and evaluated in Sections~\ref{sec:section4}–\ref{sec:section5}, see Equation~\ref{eq:saut_tenseur})
determines the distributional surface stress–energy tensor $S^{\alpha\beta}$ on $\Sigma$ via the Barrabès-Israël formalism in Poisson’s reformulation \cite{Barrabes1991,Poisson2002Reformulation} (see Section~\ref{sec:tenseur-energie-impulsion}, Equations.~\ref{eq:BI-decomp}–\ref{eq:p} for the scalars $\mu$, $j^A$ and $p$).

\section{Construction of a suitable vector basis on the null hypersurface}
\label{sec:section3}

In this section, we construct a suitable vector basis on the null hypersurface \(\Sigma\), defined by the condition \( r = \alpha \). This basis is fundamental for analyzing the geometric and physical properties of \(\Sigma\), particularly for computing the relevant tensors in the subsequent sections.

\subsection{Vector basis on \(\Sigma\)}

On the hypersurface \(\Sigma\), we define a basis comprising the following vectors:
\begin{itemize}
    \item A null vector tangent to \(\Sigma\), denoted \(k^\mu\), generating the temporal direction \(t\) \footnote{collinear with the temporal direction \(t\)},
    \item Two spacelike tangent vectors, denoted \(e_\theta^\mu\) and \(e_\phi^\mu\),
    \item A null transverse vector, denoted \(N^\mu\), which completes the basis.
\end{itemize}

\begin{figure}[htbp]
\centering
\includegraphics[width=0.8\textwidth]{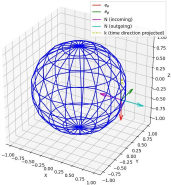} 
\caption{Schematic $3D$ representation of the vector basis on the null hypersurface \(\Sigma\) at the wormhole throat (\(r = \alpha\)), with \(\alpha = 1\) for illustrative purposes. The wireframe sphere depicts the $2D$ intrinsic geometry of \(\Sigma\), parameterized by angular coordinates \(\theta\) and \(\phi\). Vectors are shown originating from a reference point on the sphere (corresponding to Cartesian coordinates \(X=1\), \(Y=0\), \(Z=0\)). The red arrow represents the spacelike tangent vector \(e^\mu_\theta\) (pointing in the negative \(Z\)-direction). The green arrow denotes the spacelike tangent vector \(e^\mu_\phi\) (pointing in the positive \(Y\)-direction). The magenta arrow illustrates the null transverse vector \(N^\mu\) for the incoming side (associated with metric \(g^{(+)}\), directed along the negative radial direction). The cyan arrow shows \(N^\mu\) for the outgoing side (associated with metric \(g^{(-)}\), directed along the positive radial direction). The yellow dashed arrow symbolizes the null tangent vector \(k^\mu\) (or \(\ell^\mu\) in alternative notation), aligned with the temporal direction and pointing in the positive \(Z\)-direction (represented as dashed to emphasize its projected, non-spatial nature in this Euclidean visualization). This basis, as detailed in Section \ref{sec:section3}, satisfies conditions \(k^\mu k_\mu = 0\), \(N^\mu N_\mu = 0\), \(k^\mu N_\mu = 1\), \(k^\mu e^\nu_A = 0\), and \(N^\mu e^\nu_A = 0\) (where \(A = \{\theta, \phi\}\)), facilitating the computation of the transverse curvature tensor and the invariant null–shell decomposition of the surface stress-energy tensor $S^{\alpha\beta}$ using the Barrabès-Israël formalism reformulated by Poisson \cite{Poisson2002Reformulation}. Note that this is a simplified Euclidean projection for visualization. The actual geometry is embedded in the $4D$ spacetime with \(\mathcal{PT}\) symmetry.}
\label{fig:vector_basis}
\end{figure}

This vector basis is essential for calculating the surface stress-energy tensor and performing the physical decomposition in the following sections. It's illustrated in Figure~\ref{fig:vector_basis}.

\subsubsection{Null tangent vector \(k^\mu\)}

We choose \(k^\mu\) as the vector tangent to the temporal direction \( t \):
\begin{equation}\label{eq_lmu}
k^\mu = (1, 0, 0, 0).
\end{equation}

At \( r = \alpha \), the metric component \( g_{tt} = 0 \), resulting in a null norm for \(k^\mu\):
\begin{equation}
g_{\mu\nu} k^\mu k^\nu = g_{tt} = 0,
\end{equation}
confirming that \(k^\mu\) is indeed a null vector on \(\Sigma\).

\subsubsection{Spacelike tangent vectors \(e_\theta^\mu\) and \(e_\phi^\mu\)}

The angular directions \(\theta\) and \(\phi\) define vectors tangent to the \( S^2 \) sphere on \(\Sigma\):
\begin{equation}\label{eq_angular_direction_S2}
e_\theta^\mu = (0, 0, 1, 0), \quad e_\phi^\mu = (0, 0, 0, 1).
\end{equation}
Their norms are given by:
\begin{equation}
g_{\mu\nu} e_\theta^\mu e_\theta^\nu = g_{\theta\theta} = -\alpha^2 < 0,
\end{equation}
\begin{equation}
g_{\mu\nu} e_\phi^\mu e_\phi^\nu = g_{\phi\phi} = -\alpha^2 \sin^2 \theta < 0,
\end{equation}
indicating that these vectors are spacelike.

\subsubsection{Null transverse vector \(N^\mu\)}

To complete the basis, we introduce a null transverse vector \( N^\mu \) satisfying the following conditions:
\begin{equation}
k^\mu N_\mu = 1, \quad N^\mu N_\mu = 0.
\end{equation}

We set \( N^\mu = (a, b, 0, 0) \), where the coefficients \( a \) and \( b \) are determined based on the metric on each side of the hypersurface\footnote{Since a null vector cannot be normalized by itself (\(k^\mu k_\mu=0\)),
the relative scale of the null dyad \((k^\mu,N^\mu)\) is fixed by the
cross-normalization \(k^\mu N_\mu=1\) with \(N^\mu N_\mu=0\) \cite{Barrabes1991,Poisson2002Reformulation}.}.

\paragraph{Incoming side structured by the metric \( g^{(+)} \) :}

For the metric \( g^{(+)} \) as per \ref{eq_gplus_covar}, we have:
\begin{equation}
k^\mu N_\mu = g_{\mu\nu} k^\mu N^\nu = g_{tt} \cdot a + g_{tr} \cdot b = -b.
\end{equation}
The condition \( k^\mu N_\mu = 1 \) implies \( b = -1 \). \\

Furthermore,
\begin{equation}
N^\mu N_\mu = g_{\mu\nu} N^\mu N^\nu = g_{tt} a^2 + 2 g_{tr} a b + g_{rr} b^2 = 2a - 2.
\end{equation}
The condition \( N^\mu N_\mu = 0 \) gives \( 2a - 2 = 0 \), hence \( a = 1 \). \\

Thus, we obtain:
\begin{equation}\label{eq_gplus_Nmu}
N^\mu = (1, -1, 0, 0).
\end{equation}

\paragraph{Outgoing side structured by the metric \( g^{(-)} \) :}

For the metric \( g^{(-)} \), where \( g_{tr} = 1 \), the condition \( k^\mu N_\mu = 1 \) gives \( b = 1 \). Following a similar calculation, \( N^\mu N_\mu = 0 \) implies \( a = 1 \), yielding:
\begin{equation}\label{eq_gmoins_Nmu}
N^\mu = (1, 1, 0, 0).\\
\end{equation}

The adapted basis on \(\Sigma\) thus consists of \(k^\mu\), \(e_\theta^\mu\), \(e_\phi^\mu\), and \(N^\mu\), with \(N^\mu\) adjusted according to the side of the hypersurface. This construction enables an efficient decomposition of geometric tensors for the study of \(\Sigma\).

\section{Calculation of the transverse curvature tensor \( C_{ab} \)}
\label{sec:section4}

In this section, we calculate the transverse curvature tensor \( C_{ab} \) on the null hypersurface \(\Sigma\), located at \( r = \alpha \). This calculation relies on the Barrabès-Israël formalism \cite{Barrabes1991}, reformulated by Poisson \cite{Poisson2002Reformulation}, and is important for analyzing the geometric properties of \(\Sigma\).\\

The tensor \( C_{ab} \) is defined as:
\begin{equation}\label{tenseur_courbure_transverse1}
C_{ab} = e_a^\mu e_b^\nu \nabla_\mu N_\nu,
\end{equation}
where \( e_a^\mu \) are the tangent basis vectors on \(\Sigma\) and \( N^\mu \) is the null transverse vector. Since \( N^\mu \) is constant on \(\Sigma\) (i.e., \( \partial_\mu N_\nu = 0 \)), the expression reduces to:
\begin{equation}\label{tenseur_courbure_transverse2}
C_{ab} = -e_a^\mu e_b^\nu \Gamma_{\mu\nu}^\lambda N_\lambda,
\end{equation}
with \(\Gamma_{\mu\nu}^\lambda\) being the Christoffel symbols associated with the considered metric.\\

We compute the relevant components of \( C_{ab} \) for the metrics \( g^{(+)} \) (incoming side) and \( g^{(-)} \) (outgoing side), defined at \( r = \alpha \) as per \ref{eq_gplus_covar} and \ref{eq_gmoins_covar}.

\subsection{Incoming side structured by the metric \( g^{(+)} \)}

The Christoffel symbols are given by:
\begin{equation}\label{eq_christoffel}
\Gamma_{\mu\nu}^\lambda = \frac{1}{2} g^{\lambda\sigma} \left( \partial_\mu g_{\nu\sigma} + \partial_\nu g_{\mu\sigma} - \partial_\sigma g_{\mu\nu} \right).
\end{equation}

For the metric \( g^{(+)} \), as per \ref{eq_gplus_contra}, we compute the relevant components\footnote{see Appendix \ref{app:christoffel}.}:
\begin{itemize}
    \item \(\Gamma_{tt}^t\) : Since \( g_{tt} = 0 \) and the temporal derivatives are zero, but considering the non-zero terms, we obtain:
    \begin{equation}
    \Gamma_{tt}^t = \frac{1}{2\alpha}.
    \end{equation}
    
    \item \(\Gamma_{\theta\theta}^t\) : With \( g_{\theta\theta} = -r^2 \), \( \partial_r g_{\theta\theta} = -2r \), and at \( r = \alpha \), \( \partial_r g_{\theta\theta} = -2\alpha \), then:
    \begin{equation}
    \Gamma_{\theta\theta}^t = \frac{1}{2} g^{tr} (-\partial_r g_{\theta\theta}) = -\alpha.
    \end{equation}
\end{itemize}

From \ref{eq_gplus_Nmu}, the components of \( C_{ab} \) are:
\begin{itemize}
    \item \( C_{tt}^+ = -\Gamma_{tt}^t N_t \). Since \( N_t = g_{t\mu} N^\mu = g_{tr} N^r = 1 \), then:
    \begin{equation}
    C_{tt}^+ = -\Gamma_{tt}^t = -\frac{1}{2\alpha}.
    \end{equation}
    \item \( C_{\theta\theta}^+ = -\Gamma_{\theta\theta}^t N_t = \alpha \).
    \item \( C_{\phi\phi}^+ = -\Gamma_{\phi\phi}^t N_t \). 
    Since, \( g_{\phi\phi} = -r^2 \sin^2 \theta \), \( \partial_r g_{\phi\phi} = -2r \sin^2 \theta \), hence \( \Gamma_{\phi\phi}^t = -\alpha \sin^2 \theta \), yielding:
    \begin{equation}
    C_{\phi\phi}^+ = \alpha \sin^2 \theta.
    \end{equation}
\end{itemize}

Then:
\begin{equation}\label{eq_composantes_tenseur_courbure_transverse_positive}
\left\{
\begin{aligned}
    C_{tt}^+       &= -\tfrac{1}{2\alpha}, \\
    C_{\theta\theta}^+ &= +\alpha, \\
    C_{\phi\phi}^+     &= +\alpha \sin^2\theta.
\end{aligned}
\right.
\end{equation}

\subsection{Outgoing side structured by the metric \( g^{(-)} \)}

Considering the metric \( g^{(-)} \), as per \ref{eq_gmoins_contra}, the Christoffel symbols differ slightly:
\begin{equation}
\left\{
\begin{aligned}
    \Gamma_{tt}^t &= -\frac{1}{2\alpha}, \\
    \Gamma_{\theta\theta}^t &= \alpha.
\end{aligned}
\right.
\end{equation}

And from \ref{eq_gmoins_Nmu}, \( N_t = g_{tr} N^r = 1 \), we obtain:
\begin{equation}\label{eq_composantes_tenseur_courbure_transverse}
\left\{
\begin{aligned}
    C_{tt}^-       &= -\Gamma_{tt}^t = \tfrac{1}{2\alpha}, \\
    C_{\theta\theta}^- &= -\Gamma_{\theta\theta}^t = -\alpha, \\
    C_{\phi\phi}^-     &= -\Gamma_{\phi\phi}^t = -\alpha \sin^2\theta.
\end{aligned}
\right.
\end{equation}

In summary, the components \( C_{tt} \), \( C_{\theta\theta} \), and \( C_{\phi\phi} \) differ between the two sides of \(\Sigma\), reflecting the distinct properties of the metrics \( g^{(+)} \) and \( g^{(-)} \). These results will be used to determine the surface energy density and the tangential pressure through the invariant null–shell decomposition of the surface stress-energy tensor $S^{\alpha\beta}$ according to the Barrabès-Israël formalism \cite{Barrabes1991} reformulated by Poisson \cite{Poisson2002Reformulation}.

\section{Calculation of the jump in the transverse curvature tensor \( [C_{ab}] \)}
\label{sec:section5}

Across the null hypersurface \(\Sigma\) at \(r=\alpha\), the jump of the transverse curvature
\begin{equation}
[C_{ab}] \;=\; C^{+}_{ab}-C^{-}_{ab}
\label{eq:saut_tenseur}
\end{equation}
is built from the quantities \ref{tenseur_courbure_transverse1}.
From Section~\ref{sec:section4} we obtained on \(\Sigma\) (incoming \(+\) and outgoing \(-\) sides) \ref{eq_composantes_tenseur_courbure_transverse_positive} and \ref{eq_composantes_tenseur_courbure_transverse}. All mixed angular components vanish: \(C_{t\theta}=C_{t\phi}=0\).
Therefore the jump reads
\begin{equation}\label{eq_composantes_saut_courbure_transverse}
\left\{
\begin{aligned}
[C_{tt}]  &= C_{tt}^{+} - C_{tt}^{-} = -\tfrac{1}{\alpha}, \\
[C_{\theta\theta}]  &= C_{\theta\theta}^{+} - C_{\theta\theta}^{-} = 2\alpha, \\
[C_{\phi\phi}]      &= C_{\phi\phi}^{+} - C_{\phi\phi}^{-} = 2\alpha \sin^2\theta.
\end{aligned}
\right.
\end{equation}
These results will feed directly the calculation of surface energy density $\mu$ through the decomposition of the surface stress-energy tensor $S^{\alpha\beta}$ according to the Barrabès–Israël/Poisson formalism \cite{Barrabes1991,Poisson2002Reformulation}.

\section{Physical analysis of the surface stress-energy tensor at the throat}
\label{sec:tenseur-energie-impulsion}

\subsection{Null-shell decomposition}

On a null hypersurface \(\Sigma\) with null generator \(k^\alpha\) and induced two-metric
\(\sigma_{AB}\) on the space of generators (here the sphere \(S^2\) of radius \(\alpha\)),
the surface stress-energy tensor admits the invariant decomposition
\begin{equation}
S^{\alpha\beta}
= \mu\,k^\alpha k^\beta
  + j^{A}\bigl(k^\alpha e_A^\beta + e_A^\alpha k^\beta\bigr)
  + p\,\sigma^{AB} e_A^\alpha e_B^\beta ,
\label{eq:BI-decomp}
\end{equation}
where \(\mu\) is the surface energy density along the generators, \(j_A\) an angular current, and \(p\) an isotropic tangential pressure/tension.
Following Barrabès-Israël and Poisson formalism \cite{Barrabes1991,Poisson2002Reformulation}, these scalars are
\begin{align}
\mu &= -\frac{1}{8\pi}\,\sigma^{AB}[C_{AB}] ,
\label{eq:mu} \\[6pt]
j^{A} &= \frac{1}{8\pi}\,\sigma^{AB}[C_{\lambda B}] ,
\label{eq:jA} \\[6pt]
p &= -\frac{1}{8\pi}\,[C_{\lambda\lambda}]
   \;=\;\frac{1}{8\pi}\,[\kappa] ,
\label{eq:p}
\end{align}
where
\begin{subequations}\label{eq:C-definitions}
\begin{align}
C_{\lambda B} &= k^a e_B^{b} C_{ab} , \label{eq:ClambdaB}\\
C_{\lambda\lambda} &= k^a k^b C_{ab} , \label{eq:Clambdalambda}
\end{align}
\end{subequations}
and \(\sigma^{AB}\) is the inverse of the induced 2-metric \(\sigma_{AB}\) on the transverse surface. \(\kappa\) is the non-affinity defined by \(k^b\nabla_b k^a=\kappa\,k^a\), and \(C_{\lambda\lambda}=-\kappa\) follows from \(\nabla_k(k\!\cdot\!N)=0\) with \(k\!\cdot\!N=1\) on \(\Sigma\) (see Appendix \ref{sec:appendix_derivation_identity_C_lambda_lambda}). The scalars \((\mu,j^A,p)\) are invariant under the \emph{``transverse-shift’’} gauge \(N^a\!\to N^a+f\,k^a\) (with the same smooth \(f\) on both sides and \(k\!\cdot\!N=1\) preserved)\footnote{As emphasized by Poisson~\cite[p.~3 and 6]{Poisson2002Reformulation}, the degeneracy of the induced metric on a null hypersurface is not a drawback but an advantage: it allows one to define a well-behaved inverse \(\sigma^{AB}\) on the two-dimensional transverse subspace~\cite[p.~3]{Poisson2002Reformulation}. This property eliminates the trace term that appears in the non-null Israël formalism~\cite{Barrabes1991}. Furthermore, the Barrabès–Israël formalism ensures gauge invariance of the surface tensor under the freedom \(N^\mu \to N^\mu + f\,k^\mu\). The induced variation of the transverse curvature tensor, \(\delta C_{ab} = f\,k_a k_b\), lies entirely in the null direction, and thus projects to zero with \(\sigma^{AB}\). As shown explicitly in Poisson~\cite[p.~6,~(3.10)--(3.14)]{Poisson2002Reformulation}, the physical surface quantities \(\mu, j^A, p\) are therefore independent of the choice of \(N^\mu\).}.

\subsection{Application to the \(\mathcal{PT}\)-symmetric throat}

We work with the adapted basis introduced in Section~\ref{sec:section3}: \(k^\mu=(1,0,0,0)\) (null generator), \ref{eq_angular_direction_S2}, with normalization \(k\!\cdot\!N=1\).
On the sphere \(S^2\) at \(r=\alpha\), the induced two–metric and its inverse are the \(2\times2\) matrices (indices \(A,B\in\{\theta,\phi\}\))
\[
\sigma_{AB}=
\begin{pmatrix}
\alpha^{2} & 0\\[2pt]
0 & \alpha^{2}\sin^{2}\theta
\end{pmatrix},
\qquad
\sigma^{AB}=
\begin{pmatrix}
\frac{1}{\alpha^2} & 0\\[2pt]
0 & \frac{1}{\alpha^2\sin^2\theta}
\end{pmatrix}.
\]

Using the Barrabès–Israël/Poisson formalism
\ref{eq:mu} and the jumps of the transverse curvature
components from \ref{eq_composantes_saut_courbure_transverse}, we obtain
\[
\sigma^{AB}[C_{AB}]
=\sigma^{\theta\theta}[C_{\theta\theta}]
 +\sigma^{\phi\phi}[C_{\phi\phi}]
=\frac{1}{\alpha^2}(2\alpha)+\frac{1}{\alpha^2\sin^2\theta}(2\alpha\sin^2\theta)
=\frac{4}{\alpha}.
\]
Therefore,
\begin{equation}
\mu \;=\; -\frac{1}{8\pi}\,\sigma^{AB}[C_{AB}]
     \;=\; -\frac{1}{2\pi\,\alpha}\;<0 .
\label{eq:mu-final}
\end{equation}
The surface current is \ref{eq:jA} with \ref{eq:ClambdaB}. In the present adapted basis \(k^{\mu}=\partial_t^\mu\) and the only nonvanishing angular basis vectors are \(e_\theta^\mu,e_\phi^\mu\). Since \(C_{t\theta}=C_{t\phi}=0\), it follows that \(C_{\lambda\theta}=C_{\lambda\phi}=0\) and hence
\begin{equation}
j^{\theta}=j^{\phi}=0 .
\label{eq:jA-final}
\end{equation}

To obtain the pressure \(p\), we read the non-affinity \(\kappa\) from \(\nabla_k k^\mu\).
With \(k^\mu=(1,0,0,0)\), \(\nabla_k k^\mu=k^\nu\nabla_\nu k^\mu=\Gamma^\mu_{tt}\).
According to the Section \ref{sec:section4}, at \(r=\alpha\), the only non-vanishing component parallel to \(k^\mu\)
\begin{equation}
\Gamma^{t}_{tt}\Big|_{+}=\frac{1}{2\alpha},\qquad
\Gamma^{t}_{tt}\Big|_{-}=-\frac{1}{2\alpha},
\end{equation}
hence
\(
\kappa_{+}=1/(2\alpha),\ \kappa_{-}=-1/(2\alpha)
\Rightarrow [\kappa]=1/\alpha
\).
Using \ref{eq:p}, we finally get
\begin{equation}
{\ p \;=\; \frac{1}{8\pi}\,[\kappa] \;=\; \frac{1}{8\pi\,\alpha}\;>\;0\ } .
\label{eq:p-final}
\end{equation}

This positive pressure \( p \) reflects a repulsive force acting tangentially in the \( \theta \) and \( \phi \) directions on the sphere \( S^2 \). it's physically necessary to stabilize the wormhole throat against the negative surface energy density (\( \mu < 0 \)), which, without this counterbalance, would cause gravitational collapse. Moreover, \( p \) can be interpreted as a surface tension \( \tau = p \), analogous to that of a membrane or thin shell, where a positive tension keeps the structure open. This property underscores the role of exotic matter at the throat, acting as an effective fluid with repulsive effects essential for traversability.\\

Projecting \ref{eq:BI-decomp} on the sphere gives the coordinate components
\begin{equation}
{\
S_{\theta\theta}=p\,\sigma_{\theta\theta}= \frac{\alpha}{8\pi},\qquad
S_{\phi\phi}=p\,\sigma_{\phi\phi}= \frac{\alpha\sin^2\theta}{8\pi}\ } ,
\label{eq:SAB-final}
\end{equation}
while \(S_{tt}\) is not an invariantly meaningful quantity for a null shell
(physical observables are \(\mu,j^A,p\)).

\subsection{Conservation laws on the null shell}
\label{sec:Conservation-laws-on-the-null-shell}

The surface stress-energy tensor \(S^{\alpha\beta}\) satisfies the conservation equation
\begin{equation}
\nabla_{\beta} S^{\alpha\beta} = 0 \quad \text{on } \Sigma,
\end{equation}
as derived from the contracted Bianchi identities applied to the distributional form of Einstein's field equations \cite{Barrabes1991,Poisson2002Reformulation}. In the stationary and spherically symmetric configuration analyzed here, with \(j^A = 0\) and \(\partial_A \mu = \partial_A p = 0\) on the transverse 2-sphere \((S^2, \sigma_{AB})\), and given that the Levi-Civita connection \(D_A\) satisfies \(D_A \sigma^{BC} = 0\), the tangential projection of the divergence vanishes identically. This ensures that the conservation laws hold without additional constraints, consistent with the model's compatibility with Einstein's field equations (see Appendix \ref{app:conservation-null-shell}).

\section{Violation of the null energy condition (NEC)}
\label{sec:nec_violation}

In general relativity the null energy condition (NEC) requires that, for any null vector \(k^\alpha\),
\begin{equation}
T_{\alpha\beta}\,k^\alpha k^\beta \;\ge\; 0 .
\end{equation}
On a null shell \(\Sigma\) with spacetime surface stress-energy tensor \(S^{\alpha\beta}\), the distributional analogue reads
\begin{equation}
S^{\alpha\beta}\,k_\alpha k_\beta \;\ge\; 0 .
\end{equation}
Using the Poisson decomposition \ref{eq:BI-decomp} and choosing \(k^\alpha=N^\alpha\) \footnote{We work on a null hypersurface $\Sigma$ generated by the null tangent $k^\alpha$ ($k^\alpha k_\alpha=0$). Introduce an auxiliary null transverse vector $N^\alpha$ ($N^\alpha N_\alpha=0$) normalized by $k\!\cdot\!N \equiv g_{\alpha\beta}k^\alpha N^\beta=1$, and choose a basis $e_A^\alpha$ ($A=1,2$) tangent to the spacelike 2-sections $S\subset\Sigma$ with $e_A\!\cdot\!N=0$ and induced metric $\sigma_{AB}=g_{\alpha\beta}e_A^\alpha e_B^\beta$ (the dot denotes contraction with $g_{\alpha\beta}$. Angular indices are moved with $\sigma_{AB}$). With these conventions, the standard Barrabès–Israël/Poisson decomposition \ref{eq:BI-decomp} yields upon contraction $S^{\alpha\beta}N_\alpha N_\beta=\mu$, since $(k\!\cdot\!N)^2=1$ and $e_A\!\cdot\!N=0$. 
}, we obtain
\begin{equation}\label{eq_mu}
S^{\alpha\beta}N_\alpha N_\beta=\mu ,
\end{equation}
so that for a null shell the NEC reduces to the single condition
\begin{equation}
\mu \;\ge\; 0 .
\end{equation}

Therefore from \ref{eq:mu-final}, we can deduce that the NEC is explicitly violated at the throat. This encodes the presence of exotic matter required to keep the null wormhole throat open. The positive tangential pressure \ref{eq:p-final} coexists with the negative surface energy density \ref{eq:mu-final} and is fixed by the jump of the non-affinity,
consistently with the Barrabès–Israël/Poisson formalism.\\

This property is a key characteristic of traversable wormholes, as established by Morris and Thorne \cite{Morris1988}, allowing the avoidance of gravitational collapse at the throat. Hochberg and Visser~\cite{HochbergVisser1997} generalized this result, showing that for any static throat defined as a minimal-area hypersurface, the NEC is violated at least at some points on or near the throat, even without assuming spherical symmetry. Our results, based on a null hypersurface and a bimetric geometry, confirm this property within the specific framework of \( \mathcal{P}\mathcal{T} \) symmetry.\\

The exotic matter of our model, characterized by a negative surface energy density (\( \mu < 0 \)), plays a critical role in counterbalancing attractive gravitational forces. It maintains the stability of the throat at \( r = \alpha \), preventing its closure and enabling passage through \( \Sigma \) in finite time (\cite{koiran2021,koiran2024}). Thus, this property is consistent with a traversable configuration of the modified spacetime geometry, opening the door to theoretical applications such as interstellar travel.\\

While the violation of the NEC is confirmed, further studies on dynamic stability, particularly through quantum analyses, could shed light on the cosmological implications, building on \cite{Zejli2025}.

\section{Classical stability and quantum backreaction}
At the classical level considered here, the exotic matter on \(\Sigma\) does not violate any additional
conservation laws: \(\nabla_\beta S^{\alpha\beta}=0\) holds on \(\Sigma\) (see Section~\ref{sec:Conservation-laws-on-the-null-shell}), and in this stationary configuration no classical backreaction instability associated with the stress-energy along the null generators arises. Indeed, the projected conservation equations \cite{Poisson2002Reformulation} reduce to identities (see Appendix~\ref{app:conservation-null-shell}, Eq.~\ref{eq:WC-result}) when the expansion vanishes, \(\theta=0\) \footnote{Here $\mathcal{L}_k\sigma_{AB}=0$ means that the Lie derivative of the induced two–metric $\sigma_{AB}$ along the null generator $k^\mu$ vanishes: the geometry of the transverse $2$–sphere is invariant under transport along $k^\mu$ (no change with the affine/time parameter). Equivalently, $\theta=\tfrac12\,\sigma^{AB}\mathcal{L}_k\sigma_{AB}=0$. In our stationary set–up with a fixed throat at $r=\alpha$, one has $\sigma_{AB}=\alpha^2\Omega_{AB}$ independent of the parameter along $k^\mu$, hence $\mathcal{L}_k\sigma_{AB}=0$.} and the scalars \(\mu,p\) are constant.\\

To assess classical dynamical stability, one may linearize around the background and study radial perturbations, following the standard thin-shell frameworks (e.g. \cite{PoissonVisser1995} for timelike shells. The methodology must be adapted to null shells). In our case, the positive tangential pressure \(p>0\) acts as a surface tension that provides a repulsive counterbalance to collapse, suggesting stability against small radial deformations.\\

However, a complete evaluation of possible quantum backreaction effects, in the spirit of Hawking’s chronology protection conjecture \cite{Hawking1992}, which posits that quantum fluctuations may destabilize geometries admitting CTCs, requires computing the renormalized stress-energy tensor \(\langle T_{\mu\nu} \rangle_{\rm ren}\) in the full two-patch bimetric geometry. While our toy semi-classical analysis in (1+1)D (Appendix~\ref{sec:toy2D}) suggests partial flux cancellations at the throat due to \(\mathcal{PT}\)-induced frequency pairing, a rigorous (3+1)D treatment, incorporating greybody factors and Hadamard renormalization \cite{Wald1994}, is left for future work.
Such an analysis could determine whether quantum effects ultimately regularize or destabilize the traversable structure and any associated CTCs.

\section{Coherence with the \(\mathcal{PT}\)-symmetric model}
\label{sec:pt_coherence}

This wormhole model is based on a bimetric geometry defined by two conjugate metrics, \( g^{(+)} \) and \( g^{(-)} \), which exhibit a discontinuity in their first derivatives at the throat (\( r = \alpha \)). This geometric discontinuity necessarily implies the presence of a delta-distributed source localized at the throat hypersurface.

\subsection{Explicit calculation in Eddington–Finkelstein coordinates}
In ingoing/outgoing Eddington–Finkelstein coordinates adapted to the null shell at $r=\alpha$, the jump of the transverse curvature yields the surface scalars of the Barrabès–Israël/Poisson decomposition: a negative surface energy density $\mu<0$ and a positive tangential pressure $p>0$ (with $j^{A}=0$ by symmetry). This identifies the source as a lightlike membrane carrying negative energy.

\subsection{Exotic matter supporting the throat}
These values $(\mu<0,\;p>0)$ indicate an exotic matter layer required by the null junction conditions to support the throat. The positive $p$ (equivalently $[\kappa]>0$) produces an effective outward push that counterbalances the tendency to collapse, in line with lightlike brane models~\cite{guendelman2010,visser1995}.

\subsection{Effective interpretation}
The exotic matter membrane can be interpreted as an effective fluid exhibiting the following properties:
\begin{itemize}
    \item a negative surface energy density $\mu<0$,
    \item a positive isotropic tangential pressure $p>0$,
    \item dynamics associated with a lightlike brane (null surface), consistent with the analyses of Guendelman and Visser \cite{guendelman2010, visser1995}.
\end{itemize}

This repulsive effect, induced by a negative surface energy density, manifests through:
\begin{itemize}
    \item a balance of the attractive effects of the surrounding geometry,
    \item prevention of the gravitational closure of the throat,
    \item the possibility of passage through \( r = \alpha \) in finite time (\cite{koiran2021,koiran2024}),
    \item the stability of the traversability of the modified spacetime.
\end{itemize}

\subsection{Implications of the results for closed timelike curves and the role of the hypersurface \(\Sigma\)}
\label{sec:implications_CTC}
In our previous study \cite{Zejli2025}, we established that the coupling of two wormholes conforming to \(\mathcal{P} \mathcal{T}\) symmetry enables the generation of closed timelike curves (CTCs), defined as closed trajectories in spacetime satisfying modified causality conditions according to Novikov’s self-consistency principle \cite{Novikov1990}. These configurations provide a theoretical framework for exploring the properties of causality in traversable geometries. In this subsection, we establish a connection between the results of the analysis of the null hypersurface \(\Sigma\) at the wormhole throat and the characteristics of CTCs, examining the fundamental role of \(\Sigma\) in their emergence.\\

The formation of CTCs in a traversable wormhole requires two fundamental conditions: (i) the traversability of the geometry, ensuring the continuity of trajectories between the mouths and (ii) a spatial and temporal configuration connecting these mouths in a way that closes the trajectories. The current results, derived from the characterization of exotic matter at the throat via the surface stress-energy tensor \(S^{\alpha\beta}\), are directly relevant to the first condition. The violation of the null energy condition (NEC), confirmed by the surface scalar component \( \mu < 0 \) (\ref{eq:mu-final}), attests to the presence of exotic matter with negative energy density, a theoretical prerequisite for keeping the throat open and ensuring traversability, as established by Morris and Thorne \cite{Morris1988}. The absence of this matter would lead to gravitational collapse, making traversability, and consequently the formation of CTCs, impossible.\\

Additionally, the hypersurface \(\Sigma\), as a null surface defined by \( r = \alpha \), plays a critical geometric and causal role in the structure of CTCs. It serves as an interface connecting the spacetime regions described by the bimetric metrics \( g^{(+)} \) and \( g^{(-)} \), whose \(\mathcal{P} \mathcal{T}\) symmetry  ensures a structural coherence enabling non-trivial causal connections. The stability of this interface is reinforced by the positive tangential pressure, given by \ref{eq:p-final} derived from the decomposition of \( S^{\alpha\beta} \) in section \ref{sec:tenseur-energie-impulsion}. This pressure, associated with the components \( S_{\theta\theta} \) and \( S_{\phi\phi} \) given by \ref{eq:SAB-final}, counterbalances the attractive effects of the negative energy density, preventing the closure of the throat and supporting the geometry necessary for the persistence of CTCs.\\

The analysis of the null shell $\Sigma$ shows that the locally determined surface scalars $(\mu<0,\;p>0)$ are consistent with a traversable throat and with $\Sigma$ acting as a geometric stabilizer within the $\mathcal{PT}$-symmetric two-sided geometry, essential for the formation of CTCs by facilitating stable causal loops. Future investigations could quantify the nonlinear dynamics of perturbations on \(\Sigma\), building on the properties of exotic matter and the established bimetric geometry, to assess the robustness of these causal configurations in a cosmological context.

\section{Interpretation of the nature of exotic matter}
\label{sec:section9}

The $\mathcal{PT}$-symmetric wormhole relies on a violation of the null energy condition (NEC) at the throat. In the Barrabès–Israël/Poisson decomposition of the surface stress-energy tensor $S^{\alpha\beta}$ on the null shell $\Sigma$, we find a negative surface energy density $\mu<0$ and a positive tangential pressure $p>0$ (with \ref{eq_mu}). This property manifests through:
\begin{itemize}
    \item a negative {surface} energy density on the membrane ($\mu<0$),
    \item the presence of exotic matter, required by the null junction conditions to prevent local gravitational closure of the throat (the outward push is encoded in $p>0$).
\end{itemize}

This exotic matter provides the repulsive support required at the wormhole’s throat, maintaining the geometric stability of the configuration and enabling its traversability. Hochberg and Visser~\cite{HochbergVisser1997} demonstrated that the presence of exotic matter, characterized by NEC violation, is a generic property of static wormhole throats, independent of their topology or the presence of asymptotically flat regions. However, to deepen this interpretation, we propose a connection with some broader cosmological bimetric frameworks. 

\subsection{Bimetric perspective inspired by symmetry-based models}
\label{sec:janus}

The discussion below is interpretive only and does not enter our General Relativity derivations: all throat results (one–way traversability, surface energy density \(\mu<0\) and tangential pressure \(p>0\), NEC violation) follow solely from the null–shell formalism applied to the Eddington–Finkelstein metrics \ref{eq_line_element_incoming}–\ref{eq_line_element_outgoing} according to \cite{Barrabes1991,Poisson2002Reformulation}.\\

Symmetry-based cosmologies provide illustrative contexts in which a $\mathcal{PT}$-paired geometry might naturally arise. In this regard, Boyle, Finn and Turok~\cite{boyle2018cpt} proposed a \(\mathcal{CPT}\)-symmetric universe model, where the universe before the Big Bang is the \(\mathcal{CPT}\) image of the universe after, illustrating the role of time inversions in large-scale structure. This resonance suggests that our \(\mathcal{PT}\)-symmetric wormhole model could be integrated into broader cosmological frameworks, potentially to explore the arrow of time, large-scale CTCs, or fundamental symmetries of the universe. Furthermore, the \(\mathcal{PT}\)-symmetric structure of the model resonates with the bimetric Janus cosmological model of J.-P. Petit~\cite{petit2024}, which builds on Sakharov’s twin universe concept by introducing two interacting spacetime folds, each endowed with opposite arrows of time and carrying positive and negative mass sectors.\\ 

Model-independently, the throat layer is treated as a null thin shell in the Barrabès–Israël/Poisson sense. The invariants $\mu<0$ and $p>0$ are fixed by the jump of the transverse curvature and do not assume any specific microphysics \cite{Barrabes1991,Poisson2002Reformulation}. Conservative realizations include (i) a classical lightlike brane (LL-brane) sitting at the throat, with a dynamical surface tension that is on-shell negative in the explicit constructions of Guendelman et al.~\cite{guendelman2010} (see also \cite{visser1995} for the thin-shell framework), and (ii) a semiclassical vacuum–polarization layer in which the renormalized stress tensor $\langle T_{\mu\nu}\rangle_{\rm ren}$, in a Hadamard state, can locally violate the NEC \cite{BirrellDavies1982,Wald1994}. Hence, while $\mathcal{CPT}$ or Janus inspired pictures can offer broader cosmological motivation, our conclusions stand independently of them.

\subsection{Implications for traversability and stability}
Integrating this bimetric perspective enriches the analysis of the traversability of \(\mathcal{PT}\)-symmetric wormholes. Exotic matter, viewed as a consequence of interactions between sectors, doesn't merely stabilize the throat: it could also serve as a geometric bridge between the two spacetime regions connected by the wormhole. This idea suggests that wormholes might be local structures within a broader topology involving twin universes, respecting global \(\mathcal{PT}\) and \(\mathcal{CPT}\) symmetries.\\

Furthermore, this interpretation opens perspectives on the fundamental nature of exotic matter. If it arises from bimetric interactions, observational tests could be envisioned, such as studying gravitational signatures or small-scale energy fluctuations near the throat. These implications connect the properties of wormholes to broader cosmological questions, potentially addressing symmetries in the universe's large-scale structure.

\section{Predictions and observational signatures}
\label{sec:predictions}

To enhance the testability of the $\mathcal{PT}$-symmetric wormhole model, we propose several theoretical predictions grounded in the model's unique features, such as the lightlike membrane of exotic matter at the throat and the bimetric geometry with $\mathcal{PT}$ symmetry. These predictions are derived from semi-classical approximations and align with ongoing observations in gravitational wave (GW) astronomy, black hole imaging and cosmology. They distinguish our model from standard black holes or other wormhole geometries by incorporating $\mathcal{PT}$ symmetry and negative energy effects. While speculative, these signatures could be probed with current and future instruments like LIGO/Virgo, the Event Horizon Telescope (EHT) and cosmic surveys such as Euclid.

\subsection{Signatures in gravitational waves: Echoes}

\begin{figure}[t]
  \centering
  \includegraphics[width=\linewidth]{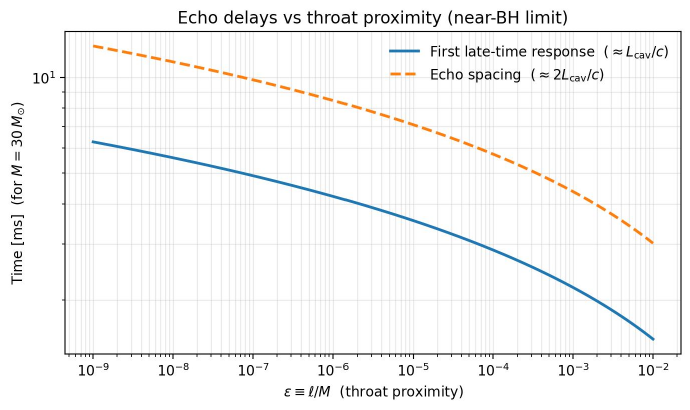}
  \caption{Echo delays as a function of the throat proximity parameter $\epsilon=\ell/M$ with $r_0=2M+\ell$, for a mass $M=30\,M_{\odot}$. The solid curve shows the arrival time of the first late-time response ($\approx L_{\rm cav}/c$) and the dashed curve shows the spacing between successive echoes ($\approx 2L_{\rm cav}/c$). The logarithmic behaviour in the near–black-hole limit ($\epsilon\ll 1$) is recovered, consistent with~\cite{Cardoso2016}.}
  \label{fig:echo_timing}
\end{figure}

The $\mathcal{PT}$-symmetric wormhole model predicts distinctive modifications to gravitational wave (GW) signals from mergers involving wormhole-like objects, e.g. a compact star plunging into the throat. Because the throat imposes nontrivial boundary conditions and the effective potential develops two barriers around the light rings, incoming GWs undergo repeated partial reflections, producing a train of late-time \emph{``echoes''} after the initial light-ring–dominated ringdown. These echoes arise from waves trapped between the photon-sphere barrier and the throat according to \cite{Cardoso2016} (see also Figure ~\ref{fig:Lcav_schematic}), with a delay time
\begin{equation}\label{eq_delay_time}
\Delta t \simeq \frac{2\,L_{\rm cav}}{c},\qquad
L_{\rm cav}=|r_*(3M)-r_*(r_0)|,
\end{equation}
which, in the near-BH limit $r_0=2M+\ell$ with $\ell\ll M$, reduces to the logarithmic scaling
\begin{equation}\label{eq_log_scaling}
\Delta t \sim -\,\frac{2M}{c}\,\ln\!\Big(\frac{\ell}{M}\Big).
\end{equation}
This logarithmic decrease with $\epsilon\equiv\ell/M$ is illustrated in Figure~\ref{fig:echo_timing}.\\

\begin{figure}[t]
  \centering
  \includegraphics[width=\linewidth]{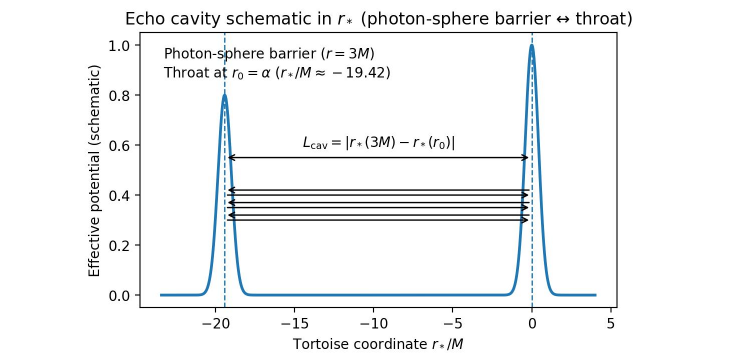}
  \caption{Conceptual schematic of the cavity in the tortoise coordinate $r_*$ between the outer barrier near the light ring ($r=3M$) and the throat located at $r_0=\alpha$. The two peaks depict the effective potential in $r_*$. The optical length $L_{\rm cav}=|r_*(3M)-r_*(r_0)|$ sets the characteristic spacing between successive echoes, $\Delta t_{\rm echo}\simeq 2L_{\rm cav}/c$. Arrows indicate multiple partial reflections that generate the late-time echo train.}
  \label{fig:Lcav_schematic_curve}
\end{figure}

This matches the generic wormhole-echo picture in which the initial ringdown is insensitive to the horizon and governed by the light ring, whereas the object's proper modes are excited only at late times. In our construction one may identify $r_0\equiv\alpha$, so that $L_{\rm cav}=|r_*(3M)-r_*(\alpha)|$ provides the natural control scale for echo delays (see Figure~\ref{fig:Lcav_schematic_curve}).\\

In the $\mathcal{PT}$-symmetric framework of this work, the conjugate metrics $g^{(+)}$ and $g^{(-)}$ map a mode of frequency $\omega$ in one sector to $-\omega$ in the other. In the unbroken $\mathcal{PT}$ regime this yields a symmetric frequency pairing. With dissipation at the exotic membrane, the pairing manifests as complex-conjugate doublets $(\omega,-\omega^*)$, providing an additional model-specific signature. Operationally, the partial \emph{``reflectivity''} entering the echo phenomenology is an effective boundary condition induced by the lightlike membrane on $\Sigma$ (see Sections~\ref{sec:tenseur-energie-impulsion}, \ref{sec:nec_violation}), rather than a fundamental mirror.

\medskip
\noindent\textit{Technical note on the delay formula.}

Working in the frequency domain, the master perturbation variable $\Psi_\ell$ satisfies the one-dimensional Schrödinger-type equation
\begin{equation}
\frac{d^2\Psi_\ell}{dr_*^2} + \bigl[\omega^2 - V_\ell(r)\bigr]\Psi_\ell = S_\ell(r;\omega),
\end{equation}
where the tortoise coordinate $r_*$ is defined by
\begin{equation}\label{eq_tortue}
\frac{dr_*}{dr} = \frac{1}{f(r)},
\end{equation}
for a static, spherically symmetric background with metric function $f(r)$. Here $V_\ell(r)$ is the effective potential and $S_\ell(r;\omega)$ the source term. Waves are partially trapped between the outer barrier near the light ring ($r_{\rm lr}\!=\!3M$ in Schwarzschild) and an inner surface at $r_0$, so the optical length of the cavity is $L_{\rm cav}=|r_*(r_{\rm lr})-r_*(r_0)|$. The spacing between successive echoes is then $\Delta t_{\rm echo}\simeq 2L_{\rm cav}/c$. Equation~(6) of~\cite{Cardoso2016} writes the first late-time response as an integral $\int dr/f(r)$, using \ref{eq_tortue} gives the compact $r_*$ form above. For $r_0=2M+\ell$ with $\ell\ll M$,
\begin{equation}
r_*(r)=r+2M\ln\!\Big(\frac{r}{2M}-1\Big)\;\Rightarrow\;
L_{\rm cav}=|r_*(3M)-r_*(2M+\ell)|
=\mathcal{O}(M)-2M\ln\!\Big(\frac{\ell}{M}\Big),
\end{equation}
hence $\Delta t_{\rm echo}\sim -\,\tfrac{2M}{c}\ln(\ell/M)$ (see Figures~\ref{fig:echo_timing} and \ref{fig:Lcav_schematic_curve}) up to additive $\mathcal{O}(M/c)$ constants that drop out of echo {spacings} (see Appendix~\ref{sec:appendix_Lcav}). In our $\mathcal{PT}$-symmetric wormhole, $r_*$ is defined so that the master equation takes the same form near $\Sigma$.  Provided $g_{tt}$ vanishes linearly at $r=r_0$ (finite surface gravity), the logarithmic behavior persists with $r_{\rm lr}$ the actual light-ring location of the effective potential.
\medskip

These features are consistent with wormhole-echo analyses showing that partial reflections produce delayed, repeating signals that can be searched for in LIGO/Virgo/KAGRA data. Conversely, the non-observation of echoes in heavy binaries can bound $L_{\rm cav}$ (and thus the throat location $r_0$ or the parameter $\alpha$) for primordial or astrophysical $\mathcal{PT}$-symmetric wormholes~\cite{Cardoso2016}.

\subsection{Optical shadows and accretion-disk images}

In our $\mathcal{PT}$-symmetric wormhole spacetime, the optical shadow and the accompanying disk emission can differ qualitatively from Kerr expectations. The absence of an event horizon together with a traversable throat at $r=\alpha$ allows null geodesics to cross between the conjugate geometries $g^{(+)}$ and $g^{(-)}$, which can yield duplicated or distorted ring-like features in the image: (i) an additional photon ring sourced by the opposite side of the throat in reflection-asymmetric wormholes, and (ii) complex \emph{“through-throat’’} views of an accretion disk located on the other side. See, respectively, double-ring/shadow predictions for asymmetric wormholes~\cite{Wielgus2020} and through-throat disk images with strong brightness distortions and multiple rings~\cite{Novikov2025}.\\

The shadow boundary is still set by unstable photon orbits (photon sphere / light ring). In our thin-shell construction, the effective potential near the throat is influenced by the shell’s stress--energy. Within our model (Section~\ref{sec:tenseur-energie-impulsion}) the exotic layer has surface energy density $\mu<0$, which modifies the local lensing environment and can induce partial reflection/transmission across the throat. In the exterior region of $g^{(+)}$ written in (ingoing/outgoing) Eddington–Finkelstein–type coordinates with a $\,\mathrm{d}r\,\mathrm{d}t\,$ cross term, we have $g_{tt}=1-\alpha/r$. For static observers\footnote{worldlines tangent to $\partial_t$}, the conserved quantity $-k_t$ and the normalization $u^t=1/\sqrt{g_{tt}}$ imply the standard gravitational redshift
\begin{equation}
\frac{\nu_\infty}{\nu_r}=\sqrt{1-\frac{\alpha}{r}},
\end{equation}
i.e. the same factor as in the Schwarzschild case because only $g_{tt}$ enters for static observers\footnote{The $g_{tr}$ term affects Doppler-like terms for moving emitters/observers, not this static redshift.}. Even without spin, the combination of gravitational and Doppler shifts can thus produce an asymmetric brightness profile in a thin disk. Moreover, in our $\mathcal{PT}$-symmetric framework the junction condition across the throat enforces a frequency pairing $\omega\mapsto -\omega$ between the two sheets (see~\cite{Zejli2025}, Section~5.1.3, Equations~(25)–(30)), which could manifest as apparent time-reversal signatures in secondary images. These last effects are model-specific predictions, not generic to wormhole shadows.\\

High-resolution VLBI observations provide empirical leverage. Current EHT images of M87* and Sgr~A* show a single, near-circular horizon-scale ring consistent with Kerr within uncertainties~\cite{EHT2019,EHT2022}. Therefore, the \emph{absence} of resolvable duplicated/through-throat ring structures at 230~GHz already imposes constraints on the parameter space of the present $\mathcal{PT}$-symmetric wormhole model (e.g., on the throat location $\alpha$ and any effective reflectance/transmittance at the membrane). A quantitative assessment, however, requires dedicated ray tracing across the throat including the junction-condition-induced reflection and transmission coefficients, as well as realistic disk/emissivity prescriptions and interstellar scattering for Sgr~A*.

\subsection{Quantum fluctuations and spectra: $\mathcal{PT}$-regulated vacuum effects}

In the $\mathcal{PT}$-symmetric framework established in~\cite{Zejli2025}, the mode amplitudes across the throat obey
$a_{-}(\omega)=a_{+}^*(-\omega)$, i.e. annihilating a mode of frequency $\omega$ on the $g^{(-)}$ side corresponds to creating a mode of frequency $-\omega$ on the $g^{(+)}$ side. The pseudo-unitary $C\mathcal{PT}$ inner product (equivalently, an $\eta$-metric) then ensures a consistent quantum dynamics with a real spectrum, despite the non-Hermitian form of the effective Hamiltonian~\cite{Kuntz2024,Zejli2025}.\\

On this basis, we conjecture that $\mathcal{PT}$ symmetry can induce a partial cancellation between conjugate frequencies $\omega$ and $-\omega$ in vacuum correlators, thereby reducing the net flux relative to the standard Hawking-like process. A fully controlled statement, however, requires an explicit semi-classical computation of the renormalized stress tensor $\langle T_{\mu\nu}\rangle$ in the present bimetric background, which calculation is detailed on Appendix \ref{sec:toy2D}. In particular, the exotic layer at the lightlike throat (with surface energy density $\mu<0$) could act as a regulator that suppresses the divergences known to arise near Cauchy horizons, yielding a locally finite state at $r=\alpha$ and a spectrally more symmetric emission between the $g^{(+)}$ and $g^{(-)}$ sectors~\cite{BirrellDavies1982,DaviesFullingUnruh1976}. This remains a testable prediction calling for dedicated QFT calculations.\\

Regarding perturbations, we expect the quasinormal-mode (QNM) spectrum to reflect $\mathcal{PT}$ symmetry. Rather than $\omega_n^*=-\omega_n$, the $\mathcal{PT}$ invariance suggests that if $\omega_n$ is a QNM frequency then the set $\{\omega_n,-\omega_n,\omega_n^*,-\omega_n^*\}$ is present, with damping rates organized in symmetric pairs. This qualitative pattern parallels $\mathcal{PT}$-symmetric scenarios in higher-derivative (quadratic) gravity, where unitarity is maintained via pseudo-Hermitian structures~\cite{Kuntz2024}. A dedicated numerical QNM analysis in the present geometry would thus provide an observational test, with possible imprints in ringdowns indicating $\mathcal{PT}$ effects.

\subsection{Cosmological implications: Contributions to acceleration and voids}

On cosmic scales, a relic population of $\mathcal{PT}$-symmetric wormholes, possibly of primordial origin, could act as an effective exotic component within symmetry-based bimetric frameworks\footnote{Our null thin-shell construction is kinematic (General Relativity and junction conditions) and does not fix the microphysics. In early-Universe settings with violent dynamics, i.e., any process that momentarily generates ultra-relativistic thin membranes, the same junction description could apply and relax to the present configuration. Quantifying production rates and stability is model-dependent and lies beyond our present scope, but will be addressed in future work.}.
The negative surface energy density at the throat ($\mu<0$) together with a positive tangential pressure (Section~\ref{sec:tenseur-energie-impulsion}) imply local repulsion, consistent with thin-shell analyses where NEC violation at the throat is generic~\cite{Lobo2004,PoissonVisser1995,visser1995}. If such objects are sufficiently abundant and randomly distributed, a coarse–grained description on FLRW domains may produce an effective fluid with $\rho_{\mathrm{eff}}+3p_{\mathrm{eff}}<0$, thereby yielding accelerated expansion without a fundamental dark-energy field. This is {formally} consistent with the standard acceleration equation
\begin{equation}
\frac{\ddot a}{a} \;=\; -\frac{4\pi G}{3}\,(\rho+3p)\,,
\end{equation}
and with the vacuum–energy interpretation of $\Lambda$ in the Einstein equations~\cite{Weinberg2008}. In that sense, one may interpret an emergent $\Lambda_{\mathrm{eff}}$ sourced by the relic wormhole population, {provided} the averaged stress–energy is approximately proportional to $g_{\mu\nu}$ on Hubble scales. An effective equation of state $w_{\mathrm{eff}}\simeq -1$ could then arise from the aggregated contribution, though a dedicated averaging calculation (e.g. in the sense of Buchert) is required to substantiate this estimate~\cite{Buchert2000}.\\

This scenario predicts large-scale underdensities (cosmic voids) generated by repulsive regions around wormhole throats and possibly asymmetric cold spots in the CMB. These signatures are testable with Euclid and DESI through void size functions, redshift-space distortions, and weak-lensing profiles~\cite{DESI2024Clustering,EuclidVoidLensing2023,EuclidVoidForecast2022,DESIVoidCatalog2024}. Observational confirmation or refutation would provide a direct and falsifiable test of the model as an alternative to $\Lambda$CDM.

\section{Conclusion}
\label{sec:section10}

In this study, we have undertaken a detailed analysis of the null hypersurface \(\Sigma\), located at the throat of a \(\mathcal{P} \mathcal{T}\)-symmetric wormhole, focusing on the characterization of the exotic matter that supports its traversability. Our results confirm the general findings of Hochberg and Visser~\cite{HochbergVisser1997}, who demonstrated that the violation of the NEC is a universal property of static throats, defined as minimal-area hypersurfaces. By employing the Barrabès–Israël/Poisson formalism \cite{Barrabes1991,Poisson2002Reformulation}, we explicitly computed the surface energy density \( \mu \) associated with this hypersurface, revealing a violation of the NEC. This violation is manifested by a negative energy density \( \mu < 0 \), substantiated through a rigorous decomposition of the components of \( S^{\alpha\beta} \) (see Sections \ref{sec:tenseur-energie-impulsion} and \ref{sec:nec_violation}). This property, consistent with the classical results of Morris and Thorne \cite{Morris1988}, constitutes an unmistakable signature of the presence of exotic matter, essential for counteracting the gravitational collapse of the throat and maintaining a traversable geometry.\\

The physical analysis of the surface stress-energy has shed light on key properties of this exotic matter. In particular, the decomposition performed in section \ref{sec:tenseur-energie-impulsion} reveals a negative surface energy density \ref{eq:mu-final} coupled with a positive tangential pressure \ref{eq:p-final}. This configuration generates a repulsive effect that stabilizes the throat, ensuring the geometric coherence necessary for traversability. Furthermore, these results integrate into the broader context of our prior work \cite{Zejli2025}, where we demonstrated that the \(\mathcal{P} \mathcal{T}\)-symmetric geometry enables the formation of closed timelike curves (CTCs). The hypersurface \(\Sigma\) plays an important role here, acting as a stabilized causal interface that facilitates non-trivial temporal connections while respecting Novikov’s self-consistency principle \cite{Novikov1990}. The positive tangential pressure reinforces this stability by counterbalancing the attractive effects of the surrounding geometry, as detailed in subsection \ref{sec:implications_CTC}.\\

Our calculations also validate the compatibility of the \(\mathcal{P} \mathcal{T}\)-symmetric model with Einstein’s field equations. The metrics $g^{(+)}$ and $g^{(-)}$ are continuous at the throat but not differentiable. The jump in their first derivatives generates a delta-supported curvature term whose coefficient is the null shell’s surface stress-energy tensor $S^{\alpha\beta}$ (with invariants $\mu, j^{A}, p$). This bimetric approach, which eliminates coordinate singularities, provides a coherent description of unidirectional traversability and supports the hypothesis of a lightlike membrane of exotic matter at the junction. Moreover, we explored a speculative interpretation of this exotic matter by drawing inspiration from symmetry-based bimetric models, such as the \(\mathcal{CPT}\)-symmetric universe of Boyle et al.~\cite{boyle2018cpt} and the Janus cosmological model of J.-P. Petit~\cite{petit2024} (subsection \ref{sec:janus}). This perspective suggests that the negative energy density could, in principle, arise from interactions between two universe sectors with opposite signed masses, an illustrative but speculative hypothesis that connects local wormhole physics to broader cosmological questions. This interpretive view is not used in any of our General Relativity derivations: all throat results follow solely from the null–shell formalism  \cite{Barrabes1991,Poisson2002Reformulation}. Conservative realizations of the exotic layer include a classical lightlike brane (LL-brane) located at the throat, with a dynamical surface tension that is on-shell negative in explicit constructions \cite{guendelman2010,visser1995}, and a semiclassical vacuum–polarization layer in which the renormalized stress tensor \(\langle T_{\mu\nu}\rangle_{\rm ren}\), in a Hadamard state, can locally violate the null energy condition (NEC) \cite{BirrellDavies1982,Wald1994}.\\

The implications of these results extend beyond merely confirming traversability. By linking the local properties of the hypersurface \(\Sigma\) to the global characteristics of spacetime, such as the formation of CTCs, this study deepens our understanding of causality in unconventional geometries. The consistency of the model with the Einstein field equations, including conservation laws on the shell and its potential extension through a bimetric interpretation enhance its theoretical significance. \\

Building on this theoretical foundation, Section~\ref{sec:predictions} translates the model into falsifiable signatures across multiple observational channels:
(i)~in gravitational waves, a thin lightlike membrane at the throat and the two-barrier potential around the light rings generically produce late-time \emph{echoes} with spacing \ref{eq_delay_time}, yielding the near-horizon logarithmic scaling \ref{eq_log_scaling} limit (Appendix~\ref{sec:appendix_Lcav}); (ii)~in horizon-scale imaging, traversability and \(\mathcal{P}\mathcal{T}\)-induced asymmetries allow duplicated or through-throat photon rings and brightness distortions absent in Kerr; (iii)~in the quantum sector, the \(\mathcal{P}\mathcal{T}\) frequency pairing \(\omega\leftrightarrow-\omega\) motivates partial cancellations in vacuum fluxes at the throat and a characteristic organization of quasinormal frequencies into symmetric (possibly complex-conjugate) doublets, as illustrated by the toy DFU analysis in Appendix~\ref{sec:toy2D}; (iv)~cosmologically, a relic population of such objects can mimic an effective \(\Lambda_{\rm eff}\) and seed voids via local repulsion, suggesting concrete tests with Euclid/DESI void statistics and weak lensing. Importantly, {non}-detections already provide constraints: the absence of resolvable echoes in heavy binaries bounds \(L_{\rm cav}\) (hence \(r_0\simeq\alpha\)), while EHT images consistent with a single Kerr-like ring restrict effective reflectance/transmittance at the membrane and the allowed throat location.\\

Several avenues emerge for further work. First, a thorough analysis of {dynamical stability} of the throat, including axial/polar perturbations and their backreaction on \(S^{\alpha\beta}\), is needed to assess robustness under realistic astrophysical conditions. In this regard, the evolving-wormhole framework of Kar and Sahdev \cite{Kar1995} is a natural template for extending our \(\mathcal{P}\mathcal{T}\)-symmetric construction beyond stationarity. Second, the \emph{GW-echo} prediction calls for targeted searches calibrated to the logarithmic spacing and to partial reflectivity at \(\Sigma\). A dedicated QNM computation in our geometry will establish precise mode spectra and couplings. Third, {ray tracing} across the throat with junction-condition, induced reflection/transmission coefficients is required to quantify imaging signatures and to confront EHT data. Fourth, a full {semi-classical QFT} treatment in \(3{+}1\)D, combining Hadamard renormalization with \(s\)-wave reduction and greybody factors, should test the \(\mathcal{P}\mathcal{T}\)-regulated flux-cancellation mechanism suggested by the \(2\)D DFU toy model (Appendix~\ref{sec:toy2D}). Finally, on cosmological scales, a controlled {averaging} (e.g.\ in the sense of Buchert) is needed to derive \(w_{\rm eff}\) and to connect the relic population to void statistics and expansion history in a way that is directly comparable with Euclid/DESI pipelines.\\

In summary, this work provides a rigorous characterization of the exotic matter supporting a \(\mathcal{P}\mathcal{T}\)-symmetric traversable throat, establishes its consistency with the Einstein field equations and ties local properties at \(\Sigma\) to global causal structure (including CTCs) and to concrete, multi-messenger observational tests. The expanded Section~\ref{sec:predictions} frames these consequences as a falsifiable program for gravitational-wave astronomy, black-hole imaging, quantum-field effects near null shells and cosmology. Together, these results strengthen the theoretical foundations of \(\mathcal{P}\mathcal{T}\)-symmetric wormholes and open a clear path to confront the model with data.

\newpage
\appendix

\section{Detailed derivation of the relevant Christoffel symbols}
\label{app:christoffel}

In this appendix, we provide a complete derivation of the Christoffel symbols relevant to the computation of the transverse curvature tensor in Section~\ref{sec:section4}, specifically \(\Gamma_{tt}^t\), \(\Gamma_{\theta\theta}^t\) and \(\Gamma_{\phi\phi}^t\), for both metrics \(g^{(+)}\) and \(g^{(-)}\). These symbols are evaluated at the throat \(r = \alpha\), but we first derive their general expressions as functions of \(r\) before substitution. By stationarity and spherical symmetry, the metric coefficients are independent of $t$ and $\phi$, and the $(t,r)$ block depends only on $r$. 
The general formula for the Christoffel symbols is given by \ref{eq_christoffel}.\\

We focus on the relevant components, noting that the inverse metric components in the \((t, r)\) sector are essential. The angular components are decoupled (\(g^{t\theta} = g^{t\phi} = 0\)), and the inverse metric is block-diagonal.

\subsection{Inverse metric components}
The metrics are given by:
\begin{itemize}
\item For \(g^{(+)}\) (incoming):
  \begin{equation}
  g_{tt}^{(+)} = 1 - \frac{\alpha}{r}, \quad g_{rr}^{(+)} = -\left(1 + \frac{\alpha}{r}\right), \quad g_{tr}^{(+)} = g_{rt}^{(+)} = -\frac{\alpha}{r},
  \end{equation}
  \begin{equation}
  g_{\theta\theta}^{(+)} = -r^2, \quad g_{\phi\phi}^{(+)} = -r^2 \sin^2 \theta.
  \end{equation}

\item For \(g^{(-)}\) (outgoing):
  \begin{equation}
  g_{tt}^{(-)} = 1 - \frac{\alpha}{r}, \quad g_{rr}^{(-)} = -\left(1 + \frac{\alpha}{r}\right), \quad g_{tr}^{(-)} = g_{rt}^{(-)} = +\frac{\alpha}{r},
  \end{equation}
  \begin{equation}
  g_{\theta\theta}^{(-)} = -r^2, \quad g_{\phi\phi}^{(-)} = -r^2 \sin^2 \theta.
  \end{equation}
\end{itemize}

The \((t, r)\) block for \(g^{(+)}\) is the matrix:
\begin{equation}
\begin{pmatrix}
1 - \frac{\alpha}{r} & -\frac{\alpha}{r} \\
-\frac{\alpha}{r} & -\left(1 + \frac{\alpha}{r}\right)
\end{pmatrix},
\end{equation}
with determinant \(-1\). The inverse is:
\begin{equation}
g_{(+)}^{tt} = 1 + \frac{\alpha}{r}, \quad g_{(+)}^{tr}  = g_{(+)}^{rt}  = -\frac{\alpha}{r}, \quad g_{(+)}^{rr}  = -\left(1 - \frac{\alpha}{r}\right).
\end{equation}

For \(g^{(-)}\), the block is:
\begin{equation}
\begin{pmatrix}
1 - \frac{\alpha}{r} & +\frac{\alpha}{r} \\
+\frac{\alpha}{r} & -\left(1 + \frac{\alpha}{r}\right)
\end{pmatrix},
\end{equation}
with determinant \(-1\). The inverse is:
\begin{equation}
g_{(-)}^{tt}  = 1 + \frac{\alpha}{r}, \quad g_{(-)}^{tr}  = g_{(-)}^{rt} = +\frac{\alpha}{r}, \quad g_{(-)}^{rr} = -\left(1 - \frac{\alpha}{r}\right).
\end{equation}

At \(r = \alpha\), these reduce to the values given in the main text: \(g^{tt} = 2\), \(g_{(+)}^{tr}  = -1\), \(g_{(-)}^{tr}  = +1\), \(g^{rr} = 0\).

\subsection{Derivation of \(\Gamma_{tt}^t\)}
For \(\Gamma_{tt}^t\):
\begin{equation}
\Gamma_{tt}^t = \frac{1}{2} g^{t\sigma} \left( 2 \partial_t g_{t\sigma} - \partial_\sigma g_{tt} \right).
\end{equation}
Since \(\partial_t = 0\), this simplifies to:
\begin{equation}
\Gamma_{tt}^t = -\frac{1}{2} g^{t\sigma} \partial_\sigma g_{tt}.
\end{equation}
Now, \(g_{tt} = 1 - \alpha/r\), so \(\partial_r g_{tt} = \alpha/r^2\), and \(\partial_\sigma g_{tt} = 0\) for \(\sigma \neq r\). Thus:
\begin{equation}
\Gamma_{tt}^t = -\frac{1}{2} g^{tr} \left( \frac{\alpha}{r^2} \right).
\end{equation}

\begin{itemize}
\item For \(g^{(+)}\): \(g_{(+)}^{tr}  = -\alpha/r\), so:
  \begin{equation}
  \Gamma_{tt}^t\Big|_{(+)} = -\frac{1}{2} \left( -\frac{\alpha}{r} \right) \left( \frac{\alpha}{r^2} \right) = \frac{1}{2} \frac{\alpha^2}{r^3}.
  \end{equation}
  At \(r = \alpha\): \(\frac{1}{2} \frac{\alpha^2}{\alpha^3} = \frac{1}{2\alpha}\).

\item For \(g^{(-)}\): \(g_{(-)}^{tr}  = +\alpha/r\), so:
  \begin{equation}
  \Gamma_{tt}^t\Big|_{(-)} = -\frac{1}{2} \left( \frac{\alpha}{r} \right) \left( \frac{\alpha}{r^2} \right) = -\frac{1}{2} \frac{\alpha^2}{r^3}.
  \end{equation}
  At \(r = \alpha\): \(-\frac{1}{2\alpha}\).
\end{itemize}

\subsection{Derivation of \(\Gamma_{\theta\theta}^t\)}
For \(\Gamma_{\theta\theta}^t\):
\begin{equation}
\Gamma_{\theta\theta}^t = \frac{1}{2} g^{t\sigma} \left( 2 \partial_\theta g_{\theta\sigma} - \partial_\sigma g_{\theta\theta} \right).
\end{equation}
Since \(\partial_\theta g_{\theta\sigma} = 0\) (metric independent of \(\theta\) in relevant components), this simplifies to:
\begin{equation}
\Gamma_{\theta\theta}^t = -\frac{1}{2} g^{t\sigma} \partial_\sigma g_{\theta\theta}.
\end{equation}
Now, \(g_{\theta\theta} = -r^2\), so \(\partial_r g_{\theta\theta} = -2r\), and \(\partial_\sigma g_{\theta\theta} = 0\) for \(\sigma \neq r\). Thus:
\begin{equation}
\Gamma_{\theta\theta}^t = -\frac{1}{2} g^{tr} (-2r) = g^{tr} r.
\end{equation}

\begin{itemize}
\item For \(g^{(+)}\): \(g_{(+)}^{tr} = -\alpha/r\), so:
  \begin{equation}
\Gamma_{\theta\theta}^t\Big|_{(+)} = \left( -\frac{\alpha}{r} \right) r = -\alpha.
  \end{equation}
  (Independent of \(r\), so same at \(r = \alpha\)).

\item For \(g^{(-)}\): \(g_{(-)}^{tr} = +\alpha/r\), so:
  \begin{equation}
\Gamma_{\theta\theta}^t\Big|_{(-)} = \left( \frac{\alpha}{r} \right) r = +\alpha.
  \end{equation}
  (Same at \(r = \alpha\)).
\end{itemize}

\subsection{Derivation of \(\Gamma_{\phi\phi}^t\)}
For \(\Gamma_{\phi\phi}^t\):
\begin{equation}
\Gamma_{\phi\phi}^t = \frac{1}{2} g^{t\sigma} \left( 2 \partial_\phi g_{\phi\sigma} - \partial_\sigma g_{\phi\phi} \right).
\end{equation}
Since \(\partial_\phi = 0\), this simplifies to:
\begin{equation}
\Gamma_{\phi\phi}^t = -\frac{1}{2} g^{t\sigma} \partial_\sigma g_{\phi\phi}.
\end{equation}
Now, \(g_{\phi\phi} = -r^2 \sin^2 \theta\), so \(\partial_r g_{\phi\phi} = -2r \sin^2 \theta\), \(\partial_\theta g_{\phi\phi} = -2 r^2 \sin \theta \cos \theta\), but since \(g^{t\theta} = 0\), only the \(r\)-derivative contributes:
\begin{equation}
\Gamma_{\phi\phi}^t = -\frac{1}{2} g^{tr} (-2r \sin^2 \theta) = g^{tr} (r \sin^2 \theta).
\end{equation}

\begin{itemize}
\item For \(g^{(+)}\): \(g_{(+)}^{tr} = -\alpha/r\), so:
  \begin{equation}
  \Gamma_{\phi\phi}^t\Big|_{(+)} = \left( -\frac{\alpha}{r} \right) (r \sin^2 \theta) = -\alpha \sin^2 \theta.
  \end{equation}
  (Same at \(r = \alpha\)).

\item For \(g^{(-)}\): \(g_{(-)}^{tr} = +\alpha/r\), so:
  \begin{equation}
  \Gamma_{\phi\phi}^t\Big|_{(-)} = \left( \frac{\alpha}{r} \right) (r \sin^2 \theta) = +\alpha \sin^2 \theta.
  \end{equation}
  (Same at \(r = \alpha\)).
\end{itemize}

\section{Derivation of the identity \(C_{\lambda\lambda}=-\kappa\) on a null hypersurface}
\label{sec:appendix_derivation_identity_C_lambda_lambda}
Let \(\Sigma\) be a null hypersurface with null generator \(k^\mu\) and auxiliary null transverse
vector \(N^\mu\) normalized by
\(k\!\cdot\!N=g_{\mu\nu}k^\mu N^\nu=1\) on \(\Sigma\).
Differentiate this normalization along \(k\) (write \(\nabla_k:=k^\mu\nabla_\mu\)):
\[
0=\nabla_k(k\!\cdot\!N)
 = (\nabla_k k)\!\cdot\!N + k\!\cdot\!(\nabla_k N),
\]
since \(k\!\cdot\!N\) is constant (=1). By definition of the non-affinity,
\(\nabla_k k=\kappa\,k\), hence
\[
(\nabla_k k)\!\cdot\!N=\kappa\,k\!\cdot\!N=\kappa.
\]
For the second term one has
\[
k\!\cdot\!(\nabla_k N)
 = k^\mu k^\nu \nabla_\mu N_\nu
 = C_{\lambda\lambda},
\]
by the definition of the transverse curvature \(C_{ab}=e_a^{\mu}e_b^{\nu}\nabla_\mu N_\nu\) projected
twice onto \(k\) (i.e. \(e_\lambda^\mu\equiv k^\mu\)).
Therefore
\[
0=\kappa + C_{\lambda\lambda}
\qquad\Longrightarrow\qquad
C_{\lambda\lambda}=-\kappa.
\]

\section{Details of the conservation law on the null shell}
\label{app:conservation-null-shell}

\subsection{Set-up and notation on {$\Sigma$}}
We work on the null hypersurface $\Sigma$ with adapted basis $\{k^\mu,\,e_A^\mu\}$, where $k^\mu$ is the null generator ($k^\mu k_\mu=0$), and $e_A^\mu$ ($A=1,2$) span the transverse two-surface $S^2$. The induced $2\times2$ metric and its inverse are
\begin{equation}
  \sigma_{AB} := g_{\mu\nu} e_A^\mu e_B^\nu, 
  \qquad \sigma^{AB}\sigma_{BC}=\delta^A_C.
\end{equation}
We use the spacetime one-forms $e_{C\alpha}:=g_{\alpha\mu}e_C^\mu$ and the mixed projectors
\begin{equation}
  e^{B}_{\beta} := \sigma^{BD}\,e_{D\beta}, 
  \qquad e_{D\beta}:=g_{\beta\mu}e_D^\mu,
\end{equation}
so that $e_{C\alpha}e_A^\alpha=\sigma_{CA}$ and $e^{B}_{\beta}e_D^\beta=\delta^B_D$.
The surface stress-energy tensor on $\Sigma$ is decomposed (following the Barrabès–Israël/Poisson formalism  \cite{Barrabes1991,Poisson2002Reformulation}) as \ref{eq:BI-decomp}.

\subsection{Distributional derivation of {$\nabla_\beta S^{\alpha\beta}=0$} on {$\Sigma$}}
Let $\tau$ be a signed distance to $\Sigma$ so that $\Sigma=\{\tau=0\}$, and write the distributional split
\begin{equation}
  T^{\alpha\beta}=T^{\alpha\beta}_{(+)}\,\Theta(\tau)+T^{\alpha\beta}_{(-)}\,\Theta(-\tau)+S^{\alpha\beta}\,\delta(\tau).
\end{equation}
The contracted Bianchi identities imply $\nabla_\beta T^{\alpha\beta}=0$ in the sense of distributions.
Testing against a smooth compactly supported scalar $\varphi$ and integrating by parts gives
\begin{equation}
  \int_\Sigma \varphi\,\nabla_\beta S^{\alpha\beta}\,\mathrm{d}\Sigma = 0 \quad \text{for all } \varphi,
\end{equation}
hence the intrinsic conservation law
\begin{equation}
\label{eq:cons-intrinsic}
  {\ \nabla_\beta S^{\alpha\beta}=0 \ \text{ on } \Sigma\ }.\footnote{When both sides of $\Sigma$ are vacuum, $T^{\alpha\beta}_\pm=0$, no external source term appears.}
\end{equation}

\subsection{Pullback to \texorpdfstring{$S^2$}{S2} and intrinsic divergence}
Define the pulled-back $(1,1)$ tensor on $S^2$ by
\begin{equation}
\label{eq:pullback-SBC}
  S^{B}_{C} := S^{\alpha\beta}\,e_{C\alpha}\,e^{B}_{\beta}.
\end{equation}
Let $D_A$ denote the Levi--Civita connection of $(S^2,\sigma_{AB})$, so that
\begin{equation}
\label{eq:compat}
  D_A \sigma_{BC}=0, \qquad
  D_A e_{C\alpha}=-\Gamma^{E}_{AC}\,e_{E\alpha}, \qquad
  D_A e^{B}_{\beta}=+\Gamma^{B}_{AD}\,e^{D}_{\beta}.
\end{equation}
The projection identity is
\begin{equation}
\label{eq:key-identity2}
  {\ D_B S^{B}_{C}
   \;=\; e_{C\alpha}\,e_{B}^{\beta}\,\nabla_\beta S^{\alpha\beta}\ } .
\end{equation}
\begin{proof} Start from $D_B(S^{\alpha\beta}e_{C\alpha}e^{B}_{\beta})$ and expand:
\begin{align}
  D_B S^{B}_{C}
   &= (D_B S^{\alpha\beta})e_{C\alpha}e^{B}_{\beta}
     + S^{\alpha\beta}(D_B e_{C\alpha})e^{B}_{\beta}
     + S^{\alpha\beta}e_{C\alpha}(D_B e^{B}_{\beta}) \nonumber\\
   &= e_{B}^{\gamma}(\nabla_\gamma S^{\alpha\beta})\,e_{C\alpha}e^{B}_{\beta}
      - \Gamma^{E}_{BC}\,S^{\alpha\beta}e_{E\alpha}e^{B}_{\beta}
      + \Gamma^{B}_{BD}\,S^{\alpha\beta}e_{C\alpha}e^{D}_{\beta} \nonumber\\
   &= e_{C\alpha}e_{B}^{\beta}\nabla_\beta S^{\alpha\beta}
      \;-\;\Gamma^{E}_{BC}\,S^{B}_{E}
      \;+\;\Gamma^{B}_{BD}\,S^{D}_{C}. \label{eq:coord-split}
\end{align}
But the last two terms are exactly the connection pieces appearing in the coordinate form
\begin{equation}
\label{eq:div-11}
  D_B S^{B}_{C}=\partial_B S^{B}_{C}
   +\Gamma^{B}_{BD}S^{D}_{C}
   -\Gamma^{E}_{BC}S^{B}_{E},
\end{equation}
so \ref{eq:coord-split} leads to \ref{eq:key-identity2} \footnote{By a “tangential projection” we compose the projector $e_{C\alpha}$ (on the free index) with $e_{B}^{\beta}$ (on the derivative index). The operator $e_{B}^{\beta}$ explicitly selects the tangential component labeled by $\beta$ along the basis direction $B$ on $S^2$. Since $e_{B}^{\gamma}e^{B}_{\beta}=q^{\gamma}_{\beta}$ is the projector onto $T(S^2)$, we obtain $e_{B}^{\gamma}(\nabla_\gamma S^{\alpha\beta})\,e_{C\alpha}e^{B}_{\beta} = e_{C\alpha}q^{\gamma}_{\beta}\nabla_\gamma S^{\alpha\beta} = e_{C\alpha}\nabla_\beta S^{\alpha\beta}$, with all non-tangential components annihilated. Finally, renaming the dummy index $\beta$ by expanding in the tangent basis via $e_{B}^{\beta}$ gives $e_{C\alpha}\nabla_\beta S^{\alpha\beta} = e_{C\alpha}\,e_{B}^{\beta}\nabla_\beta S^{\alpha\beta}$. Angular indices $A,B,C$ are raised/lowered with $\sigma^{AB}$ and $\sigma_{AB}$.}.
\end{proof}

\subsection{Tangential projection of $\nabla_\beta S^{\alpha\beta}$}
Define the tangential projection
\begin{equation}
\label{eq:WC-def}
  W_C := e_{C\alpha}\,e_{B}^{\beta}\,\nabla_\beta S^{\alpha\beta}.
\end{equation}
Using \ref{eq:key-identity2}, this is the intrinsic divergence $D_B S^{B}_{C}$ on $S^2$.
We now compute $W_C$ by inserting \ref{eq:BI-decomp} and expanding term by term.

\paragraph{Block (I): $\mu\,k^\alpha k^\beta$.}
\begin{align}
  e_{C\alpha}e_{B}^{\beta}\nabla_\beta(\mu k^\alpha k^\beta)
 &= e_{C\alpha}e_{B}^{\beta}\Big[(\partial_\beta\mu)k^\alpha k^\beta
   +\mu(\nabla_\beta k^\alpha)k^\beta
   +\mu k^\alpha\nabla_\beta k^\beta\Big] \nonumber\\
 &= 0, \qquad \text{since } e_{C\alpha}k^\alpha=0 \text{ and } e_B^\beta k^\beta=0.
\end{align}

\paragraph{Block (II): $j^A(k^\alpha e_A^\beta+e_A^\alpha k^\beta)$.}
For the two sub-terms:
\begin{align}
  e_{C\alpha}e_{B}^{\beta}\nabla_\beta(j^A k^\alpha e_A^\beta)&=0
  &&(\text{since } e_{C\alpha}k^\alpha=0),\\
  e_{C\alpha}e_{B}^{\beta}\nabla_\beta(j^A e_A^\alpha k^\beta)&=0
  &&(\text{since } e_{B}^{\beta}k^\beta=0).
\end{align}

Hence the current $j^A$ does not contribute to the {tangential} projection. Therefore, the Block (II) gives 0.

\paragraph{Block (III): $p\,\sigma^{AD}e_A^\alpha e_D^\beta$.}
Expand and use the projector identities:
\begin{align}
  T^{(3)}_C
   &= e_{C\alpha}e_{B}^{\beta}\nabla_\beta\!\big(p\,\sigma^{AD}e_A^\alpha e_D^\beta\big) \nonumber\\
   &= (D_B p)\,\sigma^{AB}\sigma_{CA}
     + p\,D_B\sigma^{AB}\,\sigma_{CA}
     + p\,\sigma^{AB}\Gamma^{E}_{BA}\,\sigma_{CE}
     + p\,\sigma_{CA}\Gamma^{B}_{BD}\,\sigma^{AD}. \label{eq:block3}
\end{align}
By metric-compatibility on $S^2$, $D_B\sigma^{AB}=0$, and the last two connection terms cancel (they are the coordinate pieces of $D_B\sigma^{B}_{C}=0$). Therefore
\begin{equation}
\label{eq:WC-result}
  {\ W_C = T^{(3)}_C = D_C p\ }.
\end{equation}

\subsection{Stationary, spherically symmetric case (used in Section~\ref{sec:Conservation-laws-on-the-null-shell})}
In the configuration considered in the main text:
\begin{equation}
  j^A=0, \qquad \partial_A p=0 \ \ (\Rightarrow D_A p=0), \qquad D_A\sigma_{BC}=0.
\end{equation}
Then \ref{eq:WC-result} yields
\begin{equation}
  e_{C\alpha}e_{B}^{\beta}\,\nabla_\beta S^{\alpha\beta}=D_C p=0
  \quad\Rightarrow\quad
  {\ D_B S^{B}_{C}=0\ \ \text{(all tangential components).} }
\end{equation}
Moreover, pulling back \ref{eq:BI-decomp} to $S^2$ gives
\begin{equation}
  S^{B}_{C}=p\,\sigma^{B}_{C} \quad\Rightarrow\quad
  D_B S^{B}_{C}=D_B(p\,\sigma^{B}_{C})
  =(D_B p)\,\sigma^{B}_{C}+p\,D_B\sigma^{B}_{C}=0,
\end{equation}
which is the same result seen directly from metric compatibility and constancy of $p$ on the sphere.

\paragraph{Remark on other projections.}
The remaining projections of $\nabla_\beta S^{\alpha\beta}$ (along $k_\alpha$ or a transverse auxiliary $N_\alpha$) are also satisfied once $S^{\alpha\beta}$ is constructed from the jump of the transverse curvature (Barrab\`es--Israël formalism): Eq.~\ref{eq:cons-intrinsic} is then a direct corollary of the contracted Bianchi identities at the distributional level. In the stationary spherical set-up, the potential kinematical scalars (expansion, shear) vanish and the scalars $(\mu,p)$ are constant along the generators, so those projections reduce to identities as well.

\section{Evaluation of the cavity length $L_{\rm cav}$}
\label{sec:appendix_Lcav}
The tortoise coordinate associated with the Schwarzschild metric of mass $M$ is
\begin{equation}
r_* = r + 2M \ln\!\left| \frac{r}{2M} - 1 \right| ,
\label{eq:tortoise}
\end{equation}
which maps the horizon $r = 2M$ to $r_* \to -\infty$. In the present configuration, the photon sphere lies at $r=3M$, while the wormhole throat is located at
\[
r_0 = 2M + \ell, \qquad \ell \ll M.
\]
The proper length of the cavity trapping the gravitational waves is thus
\begin{equation}
L_{\rm cav} = \left|\, r_*(3M) - r_*(2M+\ell) \,\right|.
\label{eq:Lcav_def}
\end{equation}

\subsection{Evaluation at $r = 3M$}
From \ref{eq:tortoise}:
\begin{equation}
r_*(3M) = 3M + 2M \ln\left( \frac{3M}{2M} - 1 \right)
= 3M - 2M \ln 2
= \mathcal{O}(M),
\end{equation}
a finite constant of order $M$.

\subsection{Evaluation at $r = 2M + \ell$}
Setting $\epsilon \equiv \ell/(2M) \ll 1$, we have:
\begin{align}
r_*(2M+\ell) 
&= (2M+\ell) + 2M \ln\!\left( \frac{\ell}{2M} \right) \nonumber \\
&= 2M + \ell + 2M \left[ \ln\!\left( \frac{\ell}{M} \right) - \ln 2 \right].
\end{align}

\subsection{Difference and simplification}
Subtracting \ref{eq:Lcav_def}, the $\pm \ln 2$ terms cancel, yielding:
\begin{equation}
L_{\rm cav} 
= \mathcal{O}(M) - 2M \ln\!\left( \frac{\ell}{M} \right),
\label{eq:Lcav_final}
\end{equation}
where $\mathcal{O}(M)$ collects all finite terms as $\ell \to 0$.\\

\begin{figure}[h!]
\centering
\begin{tikzpicture}[scale=1.0]

\draw[->] (-6,0) -- (2,0) node[right] {$r_*$};

\draw[red, thick] (-6,0) -- (-6,2);
\node[above=4pt, red] at (-6,2) {$r = 2M$ (horizon)};

\draw[blue, thick] (-5,0) -- (-5,2);
\node[above=14pt, blue] at (-5,2) {$r_0 = 2M+\ell$ (throat)};

\draw[orange, thick] (-1,0) -- (-1,2);
\node[above=4pt, orange] at (-1,2) {$r = 3M$ (photon sphere)};

\node at (-3,1.5) {Wave cavity region};

\draw [decorate,decoration={brace,amplitude=10pt,mirror}] (-5,-0.2) -- (-1,-0.2)
node[midway,below=8pt] {$L_{\rm cav}$};

\end{tikzpicture}

\caption{Schematic view of the tortoise coordinate $r_*$ showing the location of the throat $r_0=2M+\ell$, the photon sphere $r=3M$, and the effective cavity of length $L_{\rm cav}$ trapping gravitational waves.}
\label{fig:Lcav_schematic}
\end{figure}
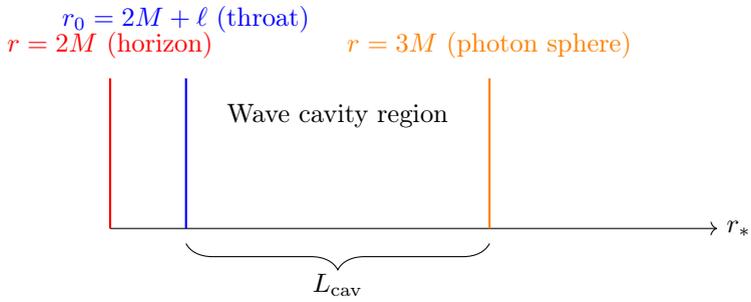

The logarithmic divergence for $\ell \ll M$ reflects the fact that, in tortoise coordinates, the throat is located far at negative $r_*$, thereby increasing the round-trip time for trapped wave packets and producing delayed \emph{``echoes''} in the late-time gravitational-wave signal (see Figure~\ref{fig:Lcav_schematic}).

\section{Toy semi-classical check in $(1+1)D$: DFU with a $\mathcal{PT}$-symmetric throat}
\label{sec:toy2D}
\subsection{Setup and DFU stress tensor (signs fixed)}

We work in the standard $(2D)$-dimensional conformal form
\begin{equation}\label{eq_metric_2D}
  \mathrm{d}s^2 = -\,C(u,v)\,\mathrm{d}u\,\mathrm{d}v,\qquad C(u,v)>0,
\end{equation}
with null coordinates $(u,v)$\footnote{Here $C(u,v)$ denotes the conformal factor in the $(1+1)$-dimensional reduction \ref{eq_metric_2D}. In the asymptotically flat region ($\mathcal{I}^+$), the geometry approaches Minkowski spacetime and $C\to 1$, so that $u$ and $v$ coincide with inertial null coordinates.}.
For any Hadamard state, point-splitting renormalization yields the DFU expressions \cite{BirrellDavies1982,DaviesFullingUnruh1976,Wald1994} (see Appendix \ref{app:2D-curvature-and-Tuv}):
\begin{align}
  \langle T_{uu}\rangle &= -\frac{1}{12\pi}\,C^{1/2}\,\partial_u^2\!\big(C^{-1/2}\big)\;+\;t_{uu}(u), \label{eq:dfu-uu}\\
  \langle T_{vv}\rangle &= -\frac{1}{12\pi}\,C^{1/2}\,\partial_v^2\!\big(C^{-1/2}\big)\;+\;t_{vv}(v), \label{eq:dfu-vv}\\
  \langle T_{uv}\rangle &= -\,\frac{1}{24\pi}\,\partial_u\partial_v\!\ln C. \label{eq:dfu-uv}
\end{align}
If the vacuum is defined by chiral reparametrizations $u\!\mapsto\!U(u)$ and $v\!\mapsto\!V(v)$, the state terms are
\begin{equation}
   \quad t_{uu}(u) \;=\; -\,\frac{1}{24\pi}\,\{U,u\}, \qquad
                 t_{vv}(v) \;=\; -\,\frac{1}{24\pi}\,\{V,v\}\quad  \label{eq:state-terms}
\end{equation}
with the Schwarzian $\{U,u\}=\dfrac{U'''}{U'}-\dfrac{3}{2}\!\left(\dfrac{U''}{U'}\right)^2$ \cite{BirrellDavies1982,ChristensenFulling1977} (see Appendix \ref{app:DFU-Schwarzian}).
For a collapsing geometry with surface gravity $\kappa$, the Unruh vacuum is specified by
\begin{equation}
  U(u)= -\,\frac{1}{\kappa}\,e^{-\kappa u}\quad\Rightarrow\quad
  \{U,u\} = -\,\frac{\kappa^2}{2}\,,
\end{equation}
hence $t_{uu} = -\tfrac{1}{24\pi}\{U,u\} = \tfrac{\kappa^2}{48\pi}$ and, with $V(v)=v$ (so $\{V,v\}=0$),
\begin{equation}
{
\langle T_{uu} \rangle \;\xrightarrow[\mathcal{I}^+]{\,C \to 1\,}\; \frac{\kappa^{2}}{48\pi},
\qquad
\langle T_{vv} \rangle \;\xrightarrow[\mathcal{I}^+]{\,C \to 1\,}\; 0
}
\label{eq:unruh-benchmark}
\end{equation}
which is the standard 2D Hawking flux benchmark \cite{BirrellDavies1982,DaviesFullingUnruh1976}.

\paragraph{Remark (4D spherical reduction):}
In spherically symmetric $4$D, the $s$-wave sector is captured by~\ref{eq:dfu-uu}--\ref{eq:dfu-uv}, with 4D flux densities obtained by the usual radial reduction (per unit solid angle, factors $\propto r^{-2}$) \cite{BirrellDavies1982,ChristensenFulling1977,DaviesFullingUnruh1976}.

\subsection{Reference states: Unruh and Hartle--Hawking (closed forms)}

\paragraph{Unruh state:}
With the choice in \ref{eq:unruh-benchmark}, and $C\!\to\!1$ asymptotically:
\begin{equation}
  { \;\langle T_{uu}\rangle_{\text{Unruh}} = \frac{\kappa^2}{48\pi}, \qquad
          \langle T_{vv}\rangle_{\text{Unruh}} = 0\; } \, .
\end{equation}

\paragraph{Hartle--Hawking state:}
Take $U(u)= - \kappa^{-1} e^{-\kappa u}$ and $V(v)= + \kappa^{-1} e^{+\kappa v}$, so
$\{U,u\}=\{V,v\}=-\kappa^2/2$ and
\begin{equation}
  { \;\langle T_{uu}\rangle_{\text{HH}} = \langle T_{vv}\rangle_{\text{HH}} = \frac{\kappa^2}{48\pi}\; }
\end{equation}
at late times / in the static patch \cite{BirrellDavies1982}.

\subsection{$\mathcal{PT}$ throat condition and frequency pairing}

In our $\mathcal{PT}$-symmetric bimetric model \cite{Zejli2025}, the throat matching reads
\begin{equation}
  \phi_-(t,r)=\phi_+^{*}(-t,r)\,,
\end{equation}
which, at the level of classical mode amplitudes, implies
\begin{equation}\label{eq_classical_amp}
  a_-(\omega) \;=\; a_+^{*}(-\omega).
\end{equation}
Upon quantization, this becomes the operator relation
\begin{equation}\label{eq_operator_amp}
  \hat a_-(\omega) \;=\; \hat a_+^{\dagger}(-\omega).
\end{equation}
We define chiral occupation numbers by
\begin{equation}
  \mathcal{N}_\pm(\omega)\;=\;\big\langle \hat a_\pm^\dagger(\omega)\,\hat a_\pm(\omega)\big\rangle.
\end{equation}
Using the canonical commutator, one finds
\begin{equation}\label{eq_canonical_commut}
  \mathcal{N}^{\rm ren}_-(\omega)\;=\;\mathcal{N}^{\rm ren}_+( -\omega)\,,
\end{equation}
and, in a stationary $\mathcal{PT}$-invariant state, $\mathcal{N}_-(\omega)=\mathcal{N}_+(\omega)$, so the $\omega$-odd parts cancel in any two-sheet sum of chiral fluxes.

\subsection{Explicit net-flux cancellation at the throat}

Define the {net outward flux across the throat} $\Sigma$ by
\begin{equation}
  \mathcal{J}^{\rm out}_\Sigma \;\equiv\; \big[\langle T_{u_+u_+}\rangle_+\big]_{\Sigma}
                                          \;-\;\big[\langle T_{v_-v_-}\rangle_-\big]_{\Sigma}\,,
\end{equation}
where the first term is the $u$-outgoing flux on the $+$ sheet and the second is the $v$-outgoing flux on the $-$ sheet (opposite orientation across $\Sigma$).

\paragraph{Assumption (local geometric symmetry):}
At the throat, the state-independent geometric pieces in \ref{eq:dfu-uu}--\ref{eq:dfu-vv} are equal on both sides (same $C$ and normal derivatives at $\Sigma$ by the junction), so they cancel in $\mathcal{J}^{\rm out}_\Sigma$. This is the standard $s$-wave approximation near a symmetric throat (4D uplift via \cite{ChristensenFulling1977}).

\paragraph{Unruh:}
On the $+$ sheet, $\langle T_{u_+u_+}\rangle_+=\kappa^2/(48\pi)$ while $\langle T_{v_+v_+}\rangle_+=0$.
Under the $\mathcal{PT}$ identification, the {outgoing} $v$-flux on the $-$ sheet equals the
{outgoing} $u$-flux on the $+$ sheet:
\begin{equation}
  \langle T_{v_-v_-}\rangle_- \;=\; \langle T_{u_+u_+}\rangle_+ \;=\; \frac{\kappa^2}{48\pi}\,,
\end{equation}
hence
\begin{equation}
  { \;\mathcal{J}^{\rm out}_\Sigma\big|_{\rm Unruh} = 0\; } \, .
\end{equation}

\paragraph{Hartle--Hawking:}
Each sheet separately has $\langle T_{u u}\rangle=\langle T_{v v}\rangle=\kappa^2/(48\pi)$, and the $\mathcal{PT}$ exchange maps $u_+$-outgoing to $v_-$-outgoing, so again
\begin{equation}
  { \;\mathcal{J}^{\rm out}_\Sigma\big|_{\rm HH} = 0\; } \, .
\end{equation}

\paragraph{Frequency-space proof (parity):}
We make explicit the hypotheses used in the parity argument at the throat $\Sigma$:
(i) the state is stationary, so that the renormalized occupations are time-independent and $\mathcal{N}_-(\omega)=\mathcal{N}_+(\omega)$;
(ii) the $\mathcal{PT}$-junction condition at $\Sigma$ implies, upon quantization, \ref{eq_operator_amp}, hence \ref{eq_canonical_commut} (see Section~5.1.3 and Appendix~A.3 of \cite{Zejli2025}).

For a given chiral sector, a schematic expression of the energy flux density reads
\begin{equation}
  \Phi_\pm\;\sim\;\int_{-\infty}^{+\infty}\!\mathrm{d}\omega\;\omega\,\mathcal{N}_\pm(\omega),
\end{equation}
where the factor $\omega$ accounts for the mode energy (overall constants are irrelevant for the parity argument).
The net two-sheet outward flux at the throat is then the difference
\begin{equation}
  \Phi_{\rm net}\;\equiv\;\int \mathrm{d}\omega\;\omega\Big[\mathcal{N}_+(\omega)-\mathcal{N}_-(\omega)\Big]
 =\int \mathrm{d}\omega\;\omega\Big[\mathcal{N}_+(\omega)-\mathcal{N}_+(-\omega)\Big],
\end{equation}
where we used $\mathcal{N}_-(\omega)=\mathcal{N}_+( -\omega)$ from (ii). Decomposing any function
$f(\omega)$ into even/odd parts,
$f(\omega)=f_{\rm even}(\omega)+f_{\rm odd}(\omega)$ with
$f_{\rm even}(\omega)=\tfrac{1}{2}\!\big(f(\omega)+f(-\omega)\big)$ and
$f_{\rm odd}(\omega)=\tfrac{1}{2}\!\big(f(\omega)-f(-\omega)\big)$, we obtain
\begin{equation}\label{eq_phi_net}
  \Phi_{\rm net} \;=\; 2\int \mathrm{d}\omega\;\omega\,\mathcal{N}_{+,\rm odd}(\omega).
\end{equation}
Hence, only the $\omega$-odd part of $\mathcal{N}_+(\omega)$ can contribute. Stationarity together with the $\mathcal{PT}$ identification at $\Sigma$ enforces the symmetry $\mathcal{N}_+(-\omega)=\mathcal{N}_+(\omega)$ in the two-sheet combination, so that $\mathcal{N}_{+,\rm odd}=0$ and the {linear} (odd) contribution vanishes identically.
Equivalently, in a low-frequency expansion,
$\mathcal{N}_+(\omega)=A+B\,\omega+C\,\omega^2+\cdots$,
\begin{equation}
  \mathcal{N}_+(\omega)-\mathcal{N}_+(-\omega)=2B\,\omega+\mathcal{O}(\omega^3)
  \;\;\Longrightarrow\;\; \Phi_{\rm net}\;\sim\;2B\int \mathrm{d}\omega\;\omega^2+\mathcal{O}(\omega^4),
\end{equation}
so the leading non-vanishing term in the two-sheet flux is {even} in $\omega$ and starts at order $\mathcal{O}(\omega^2)$.
This $\omega\leftrightarrow -\omega$ pairing is consistent with the $\omega\!\mapsto\!-\omega$ invariance of the radial equation near the throat in the slowly varying regime (Appendix~A.3 in \cite{Zejli2025}), which underlies the mode relation \ref{eq_classical_amp} (classical) and \ref{eq_operator_amp} (operators).

\begin{proof}[Frequency–parity argument at the throat]
\textbf{Assumptions:}
(A1) The state is stationary, so renormalized occupations are time-independent. 
(A2) The $\mathcal{PT}$ junction condition at $\Sigma$ implies, upon quantization, \ref{eq_operator_amp}. Hence the renormalized occupations satisfy
\begin{equation}
  \mathcal N_-^{\rm ren}(\omega)=\mathcal N_+^{\rm ren}(-\omega),
  \qquad \text{see Section~5.1.3 of \cite{Zejli2025}}.
  \label{eq:PT-occup}
\end{equation}

\textbf{Definition of net flux:}
We model the chiral energy flux density by
\begin{equation}
  \Phi_\pm \;\sim\; \int_{-\infty}^{+\infty}\! \mathrm{d}\omega\;\omega\,\mathcal N_\pm(\omega),
\end{equation}
so that the two-sheet net outward flux at the throat is
\begin{equation}
  \Phi_{\rm net}\;\equiv\;\int \mathrm{d}\omega\;\omega\Big[\mathcal N_+(\omega)-\mathcal N_-(\omega)\Big].
  \label{eq:netflux-def}
\end{equation}

\textbf{Use the $\mathcal{PT}$ identification:}
Insert \ref{eq:PT-occup} into \ref{eq:netflux-def}:
\begin{equation}
  \Phi_{\rm net}
  \;=\; \int \mathrm{d}\omega\;\omega\Big[\mathcal N_+(\omega)-\mathcal N_+(-\omega)\Big].
  \label{eq:netflux-PT}
\end{equation}

\textbf{Even/odd decomposition:}
For any function $f$, define
\(
f_{\rm even}(\omega)=\tfrac12\big[f(\omega)+f(-\omega)\big]
\)
and
\(
f_{\rm odd}(\omega)=\tfrac12\big[f(\omega)-f(-\omega)\big].
\)
Applying this to $f=\mathcal N_+$ gives the identity
\(
\mathcal N_+(\omega)-\mathcal N_+(-\omega)=2\,\mathcal N_{+,\rm odd}(\omega).
\)
Therefore \ref{eq:netflux-PT} becomes
\begin{equation}
  { \;\Phi_{\rm net} \;=\; 2\!\int \mathrm{d}\omega\;\omega\,\mathcal N_{+,\rm odd}(\omega)\; }.
  \label{eq:netflux-odd}
\end{equation}

\textbf{Consequence (cancellation of the linear/odd part):}
Under stationarity with $\mathcal{PT}$ symmetry in the two-sheet combination, $\mathcal N_+(-\omega)=\mathcal N_+(\omega)$, hence $\mathcal N_{+,\rm odd}(\omega)=0$ and the linear (odd) contribution cancels identically, i.e.\ $\Phi_{\rm net}=0$.\\

\textbf{Low-frequency corollary:}
Writing $\mathcal N_+(\omega)=A+B\,\omega+C\,\omega^2+\cdots$, one finds
$\mathcal N_{+,\rm odd}(\omega)=B\,\omega+\mathcal O(\omega^3)$, hence from \ref{eq:netflux-odd}
\(
\Phi_{\rm net}\sim 2B\!\int \mathrm{d}\omega\,\omega^2+\mathcal O(\omega^4),
\)
so the leading non-vanishing two-sheet contribution (if any asymmetry remains) is even in $\omega$ and starts at $\mathcal O(\omega^2)$.
\end{proof}

\subsection{Outcome and $4D$ uplift}
In the $(1\!+\!1)$D $s$-wave model, the above argument shows that the $\mathcal{PT}$ throat condition imposes two robust consequences:
(i) in the Unruh and Hartle--Hawking reference states, the chiral fluxes take the standard DFU closed forms,
$\langle T_{uu}\rangle_{\rm Unruh}=\kappa^2/(48\pi)$, $\langle T_{vv}\rangle_{\rm Unruh}=0$ and
$\langle T_{uu}\rangle_{\rm HH}=\langle T_{vv}\rangle_{\rm HH}=\kappa^2/(48\pi)$ (at late times / static patch);
(ii) the \emph{net} outward flux across the throat cancels exactly, because the two-sheet combination eliminates all
$\omega$-odd contributions and the state-independent geometric pieces match on both sides of $\Sigma$ (local throat symmetry).
Formally,
\begin{equation}
  {\;\Big[\langle T_{u_+u_+}\rangle\Big]_{\Sigma} \;=\; \Big[\langle T_{v_-v_-}\rangle\Big]_{\Sigma}
  \;\;\Longrightarrow\;\; \mathcal{J}^{\rm out}_\Sigma\equiv
  \Big[\langle T_{u_+u_+}\rangle\Big]_{\Sigma}-\Big[\langle T_{v_-v_-}\rangle\Big]_{\Sigma}=0\;,}
\end{equation}
for both Unruh and Hartle--Hawking states. This clarifies in which precise sense the $\omega\!\leftrightarrow\!-\omega$ pairing suppresses linear-in-frequency DFU-like terms in the two-sheet sum at the throat.\\

For a full $4D$ treatment, one must incorporate (i) the greybody factors arising from the effective scattering potential, and (ii) the backreaction on the null shell. These can be handled within the standard Hadamard point-splitting renormalization \cite{Wald1994}, combined with the $s$-wave reduction of Christensen--Fulling \cite{ChristensenFulling1977} that yields the $1/r^2$ geometric factor multiplying the $(1\!+\!1)$D DFU fluxes. The $\mathcal{PT}$-symmetric quantization is consistently formulated in a pseudo-unitary framework with a positive metric operator (``$\eta$'' or $C\mathcal{PT}$), ensuring a real spectrum and conserved inner product even for non-Hermitian effective dynamics \cite{Kuntz2024}.

\section{Conformal $(1+1)D$ geometry and the DFU mixed component $\langle T_{uv}\rangle$}
\label{app:2D-curvature-and-Tuv}

\subsection{Conformal gauge and metric components}
We work in null coordinates $(u,v)$ with the $(1+1)D$ conformal form \ref{eq_metric_2D} so that, comparing with $\mathrm{d}s^2=g_{uu}\mathrm{d}u^2+2g_{uv}\mathrm{d}u\mathrm{d}v+g_{vv}\mathrm{d}v^2$, one reads
\begin{equation}
  g_{uu}=g_{vv}=0,\qquad g_{uv}=g_{vu}=-\frac{C}{2}.
  \label{eq:A-gab}
\end{equation}
The inverse metric is purely off–diagonal:
\begin{equation}
  g^{uu}=g^{vv}=0,\qquad g^{uv}=g^{vu}=-\frac{2}{C}.
  \label{eq:A-ginv}
\end{equation}

\subsection{Christoffel symbols}
Introduce $\rho(u,v):=\tfrac12\ln C(u,v)$, so $C=e^{2\rho}$. Using
\begin{equation}
\Gamma^{a}_{bc}=\tfrac12 g^{ad}(\partial_b g_{cd}+\partial_c g_{bd}-\partial_d g_{bc}) ,
\end{equation}
and the fact that only $g_{uv}$ is non-zero, one finds the {only} non-vanishing symbols:
\begin{equation}
  \Gamma^{u}_{uu}=2\,\partial_u\rho,\qquad
  \Gamma^{v}_{vv}=2\,\partial_v\rho,
  \label{eq:A-Gamma}
\end{equation}
and all mixed ones vanish (e.g.\ $\Gamma^{u}_{uv}=\Gamma^{v}_{uv}=0$).

\subsection{Ricci tensor and scalar curvature}
From
\begin{equation}
R_{ab}=\partial_c\Gamma^{c}_{ab}-\partial_b\Gamma^{c}_{ac}
+\Gamma^{c}_{cd}\Gamma^{d}_{ab}-\Gamma^{c}_{bd}\Gamma^{d}_{ac} ,
\end{equation} with
\ref{eq:A-Gamma}, the only useful component is
\begin{equation}
  R_{uv}=-\,2\,\partial_u\partial_v\rho.
  \label{eq:A-Ruv}
\end{equation}
The scalar curvature is $R=g^{ab}R_{ab}=2\,g^{uv}R_{uv}$, hence
\begin{equation}
  R=2\!\left(-\frac{2}{C}\right)\!\left(-2\,\partial_u\partial_v\rho\right)
   = 8\,e^{-2\rho}\,\partial_u\partial_v\rho
   = 4\!\left(\frac{C_{uv}}{C^2}-\frac{C_u C_v}{C^3}\right),
  \label{eq:A-R}
\end{equation}
where subscripts denote partial derivatives\footnote{Equivalently, $R=\dfrac{4}{C}\,\partial_u\partial_v\ln C$.}.

\subsection{Trace anomaly and determination of $\langle T_{uv}\rangle$}
In two dimensions, any Hadamard state satisfies the (renormalized) trace anomaly
\begin{equation}
  \langle T\rangle := g^{ab}\langle T_{ab}\rangle = \frac{R}{24\pi}.
  \label{eq:A-anomaly}
\end{equation}
Since $g^{uu}=g^{vv}=0$, only the mixed component contributes to the trace:
\begin{equation}
  \langle T\rangle = 2\,g^{uv}\,\langle T_{uv}\rangle
  = 2\!\left(-\frac{2}{C}\right)\!\langle T_{uv}\rangle
  = -\frac{4}{C}\,\langle T_{uv}\rangle.
  \label{eq:A-trace-only-uv}
\end{equation}
Combining \ref{eq:A-anomaly} and \ref{eq:A-trace-only-uv} with \ref{eq:A-R} gives
\begin{equation}
  {\;\langle T_{uv}\rangle
   = -\,\frac{C}{4}\,\frac{R}{24\pi}
   = -\,\frac{1}{24\pi}\,\partial_u\partial_v\ln C\;}
  \label{eq:A-Tuv}
\end{equation}
which is precisely the DFU mixed component in conformal gauge. This formula is the state-{independent} geometric piece. State dependence only enters $\langle T_{uu}\rangle$ and $\langle T_{vv}\rangle$ through chiral reparametrizations (Schwarzian terms), whereas $\langle T_{uv}\rangle$ is fully fixed by the anomaly and conservation. 

\subsection{Consistency checks and remarks}
\paragraph{(i) Flat limit:} If $C\to 1$\footnote{Minkowski in null coordinates}, then $\partial_u\partial_v\ln C=0$ and \ref{eq:A-Tuv} gives $\langle T_{uv}\rangle=0$, as expected.

\paragraph{(ii) Conservation:} If we assemble the three components that constitute the complete DFU stress tensor, equations~\ref{eq:dfu-uu} and \ref{eq:dfu-vv} for $\langle T_{uu}\rangle$ and $\langle T_{vv}\rangle$, together with equation~\ref{eq:A-Tuv} for $\langle T_{uv}\rangle$, the resulting tensor satisfies the covariant conservation law $\nabla^{a}\langle T_{ab}\rangle=0$ for any choice of the chiral reparametrizations $U(u)$ and $V(v)$ defining the Hadamard state. The state-dependent terms $t_{uu}(u)$ and $t_{vv}(v)$ are precisely those required to preserve covariance under these reparametrizations. A derivation of the state terms from the identity relating DFU operators in $(U,V)$ and $(u,v)$ coordinates, and the appearance of the Schwarzian,
is given in Appendix~\ref{app:DFU-Schwarzian}.

\paragraph{(iii) $4$D $s$-wave uplift:} In spherically symmetric $4$D, the $(1{+}1)$D
DFU fluxes uplift to $4$D flux densities via the usual radial reduction by a factor $1/r^2$ (per unit solid angle) \cite{ChristensenFulling1977}.

\section{State-dependent DFU terms and the Schwarzian}
\label{app:DFU-Schwarzian}

We work in $(1+1)$ dimensions with the conformal gauge \ref{eq_metric_2D}, and adopt the DFU renormalized expressions \ref{eq:dfu-uu} and \ref{eq:dfu-vv} for any Hadamard state \cite{BirrellDavies1982,DaviesFullingUnruh1976,Wald1994}.

\subsection{Setup of the vacuum and coordinate reparametrizations}
We define the vacuum by choosing null reparametrizations
\begin{equation}
  U=U(u),\qquad V=V(v),\qquad U'(u)>0,\;V'(v)>0,
  \label{eq:reparam}
\end{equation}
such that in $(U,V)$ coordinates the outgoing/ingoing fluxes vanish:
\begin{equation}
  \langle T_{UU}\rangle=\langle T_{VV}\rangle=0.
  \label{eq:vac}
\end{equation}
Under \ref{eq:reparam} the metric remains conformal with
\begin{equation}
  \tilde C(U,V)=\frac{C(u,v)}{U'(u)\,V'(v)},\qquad
  \partial_U=\frac{1}{U'}\,\partial_u,\qquad
  \partial_V=\frac{1}{V'}\,\partial_v .
  \label{eq:Ctilde-and-derivs}
\end{equation}
Tensorial transformation gives $T_{uu}=(U')^2T_{UU}$, $T_{vv}=(V')^2T_{VV}$.\\

From \ref{eq:vac} and the DFU form in $(U,V)$ we have
\begin{equation}
  \langle T_{UU}\rangle=-\frac{1}{12\pi}\,\tilde C^{1/2}\,\partial_U^2(\tilde C^{-1/2}),\qquad
  \langle T_{VV}\rangle=-\frac{1}{12\pi}\,\tilde C^{1/2}\,\partial_V^2(\tilde C^{-1/2}).
\end{equation}
Therefore, in $(u,v)$ coordinates,
\begin{equation}
  \langle T_{uu}\rangle= -\frac{(U')^{2}}{12\pi}\,\tilde C^{1/2}\,\partial_U^2(\tilde C^{-1/2}),\qquad
  \langle T_{vv}\rangle= -\frac{(V')^{2}}{12\pi}\,\tilde C^{1/2}\,\partial_V^2(\tilde C^{-1/2}).
  \label{eq:Tuu-from-U}
\end{equation}
Comparing \ref{eq:Tuu-from-U} with \ref{eq:dfu-uu} and \ref{eq:dfu-vv} shows that the state-dependent pieces $t_{uu}$ and $t_{vv}$ are entirely encoded in the difference between the two DFU operators built from $C$ and $\tilde C$.

\subsection{Key identity}
Set
\begin{equation}
  F(u):=(U')^{1/2}C^{-1/2}(u,v),\qquad
  \text{so that } \quad \tilde C^{-1/2}=V'^{1/2}F,
\end{equation}
and note that $V'$ is $u$–independent. Using \ref{eq:Ctilde-and-derivs},
\begin{equation}
  (U')^{2}\tilde C^{1/2}\,\partial_U^2(\tilde C^{-1/2})
  =(U')^{3/2}C^{1/2}\,\partial_U^2F
  =C^{1/2}\!\left[-\frac{U''}{(U')^{1/2}}\partial_uF+\frac{1}{(U')^{1/2}}\partial_u^2F\right].
  \label{eq:block-left}
\end{equation}
Direct differentiation gives
\begin{equation}
  \partial_uF=\frac{1}{2}\frac{U''}{U'}(U')^{1/2}C^{-1/2}-\frac{1}{2}(U')^{1/2}C^{-3/2}C_u,
\end{equation}
\begin{equation}
\begin{split}
  \partial_u^2F&=\frac{1}{2}\!\left(\frac{U'''}{U'}-\Big(\frac{U''}{U'}\Big)^2\right)(U')^{1/2}C^{-1/2}
  +\frac{1}{4}\!\left(\frac{U''}{U'}\right)^{\!2}(U')^{1/2}C^{-1/2} \\
  &\quad -\frac{1}{4}\frac{U''}{U'}(U')^{1/2}C^{-3/2}C_u
  -\frac{1}{2}(U')^{1/2}\!\left[-\frac{3}{2}C^{-5/2}C_u^2+\frac{1}{2}C^{-3/2}C_{uu}\right].
  \end{split}
\end{equation}
Moreover
\begin{equation}
  C^{1/2}\partial_u^2(C^{-1/2})=\frac{3}{4}C^{-2}C_u^2-\frac{1}{2}C^{-1}C_{uu}.
\end{equation}
Substituting the last three formulae into \ref{eq:block-left}, and subtracting $C^{1/2}\partial_u^2(C^{-1/2})$, all terms containing $C,\,C_u,\,C_{uu}$ cancel identically, leaving
\begin{equation}
  (U')^{2}\tilde C^{1/2}\,\partial_U^2(\tilde C^{-1/2})
  -C^{1/2}\partial_u^2(C^{-1/2})
  =\frac{1}{2}\left[\frac{U'''}{U'}-\frac{3}{2}\left(\frac{U''}{U'}\right)^{2}\right].
  \label{eq:key-identity}
\end{equation}
The bracket is the Schwarzian derivative
\begin{equation}
  \{U,u\}:=\frac{U'''}{U'}-\frac{3}{2}\left(\frac{U''}{U'}\right)^{2}.
\end{equation}

\subsection{State terms}
Inserting \ref{eq:key-identity} into \ref{eq:Tuu-from-U} and comparing with \ref{eq:dfu-uu} and \ref{eq:dfu-vv} yields
\begin{equation}
  t_{uu}(u)=-\frac{1}{24\pi}\,\{U,u\},\qquad
  t_{vv}(v)=-\frac{1}{24\pi}\,\{V,v\}.
  \label{eq:state-terms}
\end{equation}

\newpage
\addcontentsline{toc}{section}{References}
\providecommand{\noopsort}[1]{}\providecommand{\singleletter}[1]{#1}%

\end{document}